%% file: 2025_DDBC_p3_arxiv.tex
\documentclass[twocolumn]{aastex701}
\usepackage{newunicodechar,graphicx} % to get the hawaiian okina
\usepackage{ulem}
\usepackage{amsmath}
\input{macros}

\begin{document}

\title{Simulating the photospheric to coronal plasma using magnetohydrodynamic characteristics III: validation including gravity, flux emergence, and an eruption}

\correspondingauthor{Lucas A. Tarr}
\author[0000-0002-8259-8303,gname=Lucas,sname=Tarr]{Lucas A. Tarr}
\email[show]{ltarr@nso.edu}
\affiliation{National Solar Observatory, 22 Ohi\okina{}a Ku St, Makawao, HI, 96768}

\author[0000-0001-7322-5401,gname=N. Dylan,sname=Kee]{N. Dylan Kee}
\affil{NASA Goddard Space Flight Center, 8800 Greenbelt Rd, Greenbelt, MD 20771}
\email{nathaniel.d.kee@nasa.gov}

\author[0000-0002-6936-9995,gname=James,sname= Leake]{James E. Leake}
\affil{NASA Goddard Space Flight Center, 8800 Greenbelt Rd, Greenbelt, MD 20771}
\email{james.e.leake@nasa.gov}

\author[0000-0002-4459-7510,gname=Mark,sname= Linton]{Mark G. Linton}
\affil{U.S. Naval Research Laboratory, 4555 Overlook Ave SW, Washington, DC 20375}
\email{mark.g.linton.civ@us.navy.mil}

\author[0000-0003-1522-4632,gname=Peter,sname= Schuck]{Peter W. Schuck}
\affil{NASA Goddard Space Flight Center, 8800 Greenbelt Rd, Greenbelt, MD 20771}
\email{peter.schuck@nasa.gov}

\begin{abstract}
Solar eruptions arise from instabilities or loss of equilibria in the solar atmosphere, but routinely inferring the precise magnetic and plasma properties that lead to eruptions is not currently practical using synoptic solar observations.
Data driven simulations offer an appealing alternative.
We test our boundary data-driven magnetohydrodynamic (MHD) approach, based on the method of characteristics, on a simulation that includes full MHD, a stratified atmosphere, and the emergence of a model solar magnetic active region, from the photosphere upwards.
The driven simulation is tested against a larger, ab initio ``Ground Truth'' simulation that extends downward into the convection zone.
Our driven simulation accurately reproduces the dynamic emergence of the active region above the photosphere, the formation of key topological features throughout the corona, and the subsequent eruption of mass and magnetic field.
The total emerged energy matches to better than one percent, the ratio of emerged to eruptive energy is $\approx2\%$, and the actual values of each energy term agree to within $10\%$ between the two cases.
Crucially, the data injection cadence, when properly scaled, matches the cadence of synoptic observations of the Sun's surface magnetic field, and is three to four orders of magnitude longer than the inherent CFL time step of the simulations.
The stability of the code and fidelity of the results over an entire active region lifetime, from emergence to eruption, strongly suggests that our method will produce reliable results when driven using solar synoptic observations from existing and anticipated ground and spaced based observatories.
\end{abstract}


\section{Introduction} \label{sec:intro}
This is the third paper in a series describing the systematic application of the method of characteristics \citep{Courant:1953, Morse:1953a, Hedstrom:1979, Thompson:1987a,Thompson:1990} to boundary conditions for magnetohydrodynamics (MHD) simulations,
with the goal of studying the evolution of magnetic active regions on the Sun.
In \citet{Tarr:2024}, hereafter \citetalias{Tarr:2024}, we laid out the mathematical description of general boundary conditions in terms of the characteristic formulation of MHD, its application to (boundary) data driven simulations in particular, and the numerical implementation and testing of our data driving method on a toy problem.
The numerical boundary condition framework implemented in \citetalias{Tarr:2024} is called \charc{}, and was initially developed as a module extending the \lare{} MHD code \citep{Arber:2001}; a standalone module capable of being called by most MHD codes is in development.
The module is general, and any physically admissible boundary conditions can be implemented using it.
For example, in \citet{Kee:2025}, hereafter \citetalias{Kee:2025}, we analyzed the performance of two common versions of so-called ``non-reflecting boundary conditions'' \citep[][]{Hedstrom:1979, Thompson:1990, Cimino:2016} on a suite of test problems.
Data driven boundary conditions and the various flavors of nonreflecting boundary conditions are all types of open boundary conditions, and are conceptually closely related.
Nonreflecting boundary conditions attempt to accurately model information leaving a domain and data driven boundary conditions attempt to use physical variables inferred from remote (and potentially in situ) observations to accurately model information entering a domain.
Both are derived from an accurate representation of how information propagates though an MHD system, with the characteristics being the natural mathematical description of those processes anywhere in a volume.
An open boundary condition then specifies the bidirectional interaction, in terms of the flow of information, between a given domain of interest and the external universe.
This third paper in our series refocuses attention back to the data driving problem by testing a numerical setup that is much more representative of conditions on the Sun than the toy models we have previously considered.

Our motivation in developing data driven boundary conditions is to accurately reproduce key features of the low solar atmosphere, between the photosphere and the Alfv\'en surface, in order to understand the origin of solar flares and coronal mass ejections.
Essentially, we want to model the evolution of solar active regions from their emergence through the photosphere to their eruption in the corona above.
We focus on data driving using photospheric data for two main reasons. 
First, within the photosphere, the inference of the state of the plasma from remote observations of polarized emission of atomic spectral lines is a much more tractable problem than in any other layer of the solar atmosphere; and second, the solar physics community has long-running, publicly available synoptic observations at high enough spatial resolution and temporal cadence to make boundary data driving feasible \citep{Harvey:1996, Scherrer:1995,Scherrer:2012}.
Previous theoretical work and observational evidence suggests the presence of twisted flux ropes, magnetic null points and separators, thin current layers, and a complex distribution of plasma pressure within solar active regions, all of which combine to create the instabilities or loss of equilibria that ultimately precipitate eruptions \citep{Priest:2014}.
The appeal of a data-driven approach is that, if properly implemented, these features of the solar atmosphere will arise dynamically and self-consistently within the driven simulation, with a minimum of ad-hoc assumptions.

Numerous data driving methods for solar active regions have been published very recently, including \citet{Fan:2022, Afanasyev:2023,Chen:2023, Inoue:2023,Guo:2024a,Fan:2024,Liu:2024,Wagner:2024,Chen:2025}; see also relevant review articles by \citet{Jiang:2021}, \citet{Guo:2024b}, and \citet{Schmieder:2024}.
Not all of these studies are directly comparable, given the relative novelty of the field, the rapid progress by different groups on different techniques, and the use of the term ``data driving'' to (justifiably) apply to all of them.
Most of these involve applying a succession of different techniques to various stages of the emergence--to--eruption lifetime of an active region, some ad hoc, which collectively can be described as data driven or data constrained approaches. 
The steps are varied and can include: various magnetic field extrapolations, magnetofrictional relaxation, ad hoc emergence using semi-analytic flux tube models, randomized electric field driving, switching from (quasi-driven) magnetofrictional to (perhaps line-tied) MHD relaxation, and so on.
The breadth and variety of methods above are not of particular interest here, so we will not describe them in detail.
Instead, we note several features that distinguish our method from the others. 

In this paper, we define a ``data driven simulation'' as an ongoing, dynamic simulation where the lower boundary conditions for the primitive variables\footnote{The specifics of how the primitive variables are set vary depending on the numerical scheme, e.g., a finite volume scheme might set them in terms of fluxes of conserved quantities.  Mathematically, all schemes must be equivalent to using Dirichlet, Neumann, or Robin boundary conditions on the primitive variables.} are updated at each Courant condition-limited time step using some form of temporal interpolation (perhaps with additional modifications) between longer-spaced observation times.
With that in mind, our approach is relatively straight-forward, requiring a minimum of steps and assumptions: temporally interpolated observational data are used as directly as possible, and only modified to be consistent with known physical laws.
While there can be some choice in how to adhere to physical laws given a time series of data that includes observational errors or uncertainties, that is the only choice we allow ourselves to make.
All specifics of the point-to-point and time-to-time implementation of those choices are determined using an optimization technique.
The only ad hoc part of our method is the choice of the initial condition.

In more detail, our method is designed to incorporate observational inference of the entire MHD state vector---the density $\rho$, internal energy density $\epsilon$, velocity $\vect{v}$, and magnetic field $\vect{B}$---at each location and at each time, into the driving scheme, utilizing an optimization approach capable of handling missing, noisy, and biased data, which are always present in real observations.
The mathematical and numerical framework we have developed in \citetalias{Tarr:2024}, including the optimization step, is described entirely in terms of the characteristics of the MHD system.
That choice forces the evolution of the driven boundary to be consistent with the MHD equations (up to the fidelity of the numerical scheme), a requirement not guaranteed by other data driven approaches.
A corollary is that our simulations can and will halt if they reach a point where there is no physically allowed evolution of the boundary that moves the system closer toward the observations.
Reaching such a condition is a strong indication that the solution in the volume away from the boundary has departed so far from reality that the results can no longer be trusted.
We consider this a strength of our method.

Beyond the differences in methodology, the work we present here uses a validation strategy that differs somewhat from other efforts.
Of the studies listed above, \citet{Chen:2023} performs the most similar test to ours, beginning with a ground truth simulation in a larger domain, then running their data driving method and comparing the results in a smaller domain, all in framework of full MHD (and in their case, radiative MHD).
While many differences exist between both the methods and the validation tests in either case, we find three that are particularly important. 
First, their method uses a lower boundary that is closed to vertical mass flux, and it is therefore also closed to magnetic flux for the approximately ideal evolution of the photospheric boundary. 
At the same time, they do force an evolving magnetic field that is associated with flux emergence, so there is some inconsistency in the description and solution of the problem in their case.
Second, their method attempts to use and match only the evolution of the vertical component of the magnetic field on the boundary (by modifying the horizontal components as necessary), whereas our method can---and in this case, does---utilize the entire MHD state vector. 
Third, their test case focuses on the emergence of a parasitic bipolar region that is essentially a perturbation to a much larger, stable active region structure, whereas we study the emergence of a full active region from whole cloth: our test temporally covers the entire duration of active region emergence, including an eruption, during which time the emerged field becomes by far the dominant coronal structure in the simulation.

In a recent preprint, \citet{Chen:2025} report that they apply a version of their method to observational data of NOAA AR 11158, using one of the hybrid approaches mentioned above that consists of an extended zero-$\beta$ simulation that is later expanded to full (radiative) MHD to model the eruption of a flux rope.
This new test demonstrates that their method is also capable of handling cases where the emerging field becomes the dominant feature of the corona.
However, it still includes the inconsistency of a lower boundary that is closed to vertical mass flux while still allowing emerging magnetic field. 
Importantly, they also apply an ad hoc rotation to every vertical flux element, which serves as the primary mechanism to inject free magnetic energy (i.e., electric currents) into the volume.
Control experiments with no rotation, or approximately half the rotation used in the final simulation, do not produce eruptions.
This is interesting because it implies that it is twist on individual, pixel scale flux elements, and not the bulk motion and associated large-scale currents of the flux emergence process itself, that ultimately lead to the eruptions.

\citet{Inoue:2023} also compare their data driving method to a ground truth simulation, but in a way that is substantially different from either the present case or that in \citet{Chen:2023}. 
For \citet{Inoue:2023}, the extent of both the ground truth and driven simulations are the same.
Their ground truth simulation starts with a preexisting potential field, to which a velocity driver is applied in the lower boundary to twist up the field and launch an eruption, which is a type of ad hoc driving.
The ``data driven'' simulation then takes the ad hoc driven lower boundary of the ground truth simulation and attempts to recreate that driving by different means that could, in principle, be applied to observational data.
It also does not include flux emergence, and indeed, the distribution of vertical flux in the lower boundary is fixed throughout the simulation.
So, while it is an interesting test, it is quite different in spirit from what we undertake here.
As far as we are aware, we are the first to perform a data driven MHD simulation of an active region, from pre-emergence to eruption, using a single technique with no free parameters other than the initial condition (which itself is a major open avenue of research for such simulations; see \citet{Barnes:2024}).

The present work is a rigorous validation exercise as advocated by \citet{Leake:2017} and undertaken in \citetalias{Tarr:2024} and \citetalias{Kee:2025}.
It proceeds as follows:
\begin{enumerate}
    {\item Run a ``Ground Truth'' (\GTsim{}) simulation, spanning the convection zone to the corona, to serve as a reference.}
    {\item Extract a time series of synthetic observations from the interior of the \GTsim{} simulation at a single, photospheric height.}
    {\item Use the extracted synthetic observations as input to our data driven boundary conditions for the Data Driven (\DDsim{}) simulation, spanning the photosphere to corona.}
    {\item Quantify how well the \DDsim{} simulation matches the \GTsim{} simulation in the overlapping photosphere-to-corona domain.}
\end{enumerate}
Compared to our previous work, this new test uses a \GTsim{} simulation with a more comprehensive suite of physical characteristics and interesting dynamics applicable to the Sun: it includes gravity, a stratified solar-like atmosphere, magnetic flux emergence from a buoyantly-rising twisted flux rope, topological changes associated with magnetic reconnection, and, finally, an eruption that rapidly carries mass and magnetic flux into the upper corona.
This validation study therefore captures many of the features of interest to us in the solar corona and gives us confidence that the method can produce reliable results when applied to real observational data.

The rest of the paper is organized as follows.  
In \autoref{sec:GTsim} we describe relevant aspects of the \GTsim{} simulation, which was previously reported on in \citet{Leake:2022}.
In \autoref{sec:syntheticBC} we describe the synthetic observations, extracted from the \GTsim{} simulation, that will serve as the inputs to our data driving method.
In \autoref{sec:DDsim} we describe the numerical setup of the DD simulation, including details that distinguish it from our previous work.
In \autoref{sec:results} we present our results, comparing the DD simulation to the \GTsim{} solution across a wide variety of metrics, including the mass, energy, and Poynting fluxes, the timing of various events, and the presence of key topological features of the emerged field.
We further discuss the results and conclude in \autoref{sec:discussion}.

\section{Ground truth simulation} \label{sec:GTsim}

\begin{figure*}
    \centering
    \includegraphics[width=1.0\linewidth]{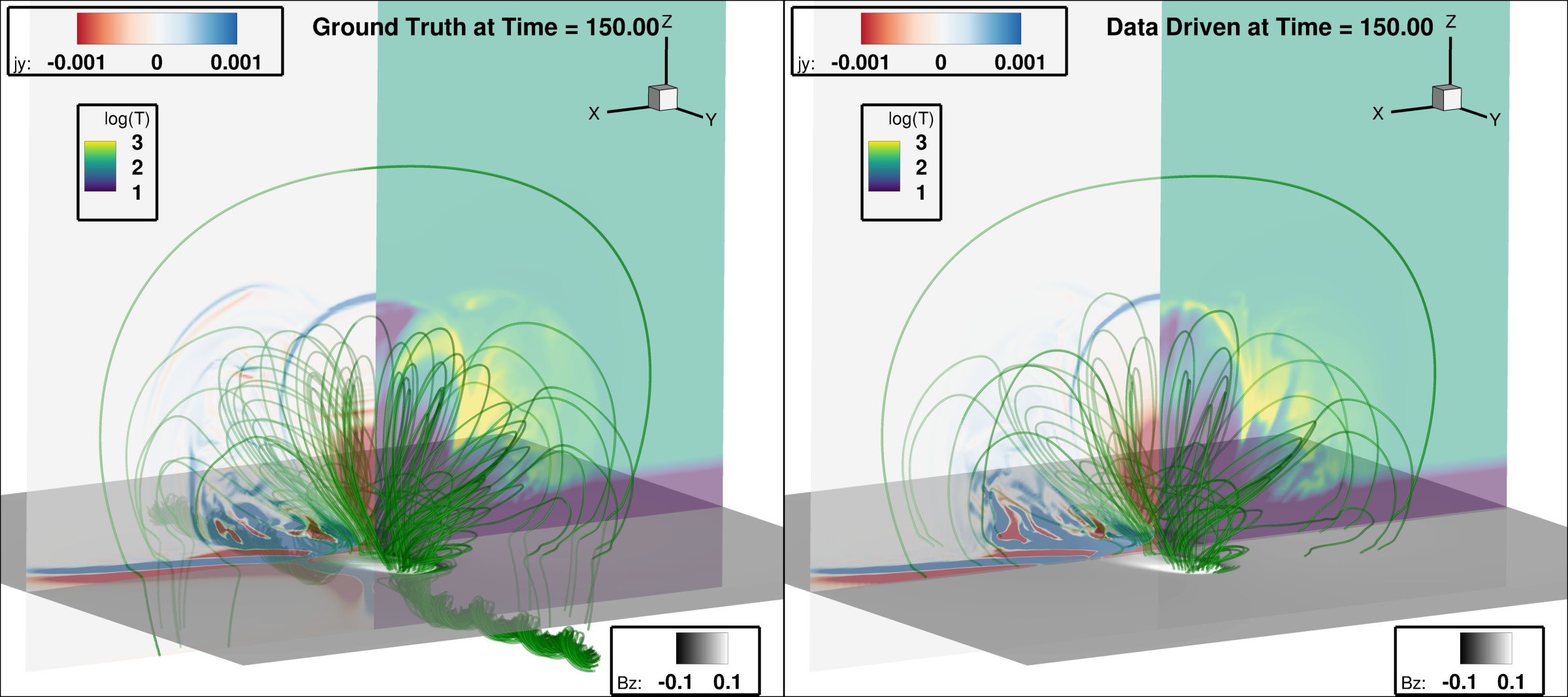}
    \caption{Perspective comparison between the \GTsim{} simulation (left) and DD simulation (right) at time $t=150$.  The spatial extent of the image is the same in both panels and extends vertically into the convection zone, which is below the numerical domain of the \DDsim{} simulation.  The flux tube in the convection zone is therefore visible in the left panel but not the right.  The same features are displayed for each simulation, consisting of the vertical magnetic field in the $z=0$ plane (grayscale); the current density $j_y$ along the primary axial direction of the flux tube in the $y=0$ plane (Red-blue scale); the temperature in the $y=0$ plane (purple-yellow scale); and selected magnetic field lines in both the emerged flux rope and background magnetic arcade (green lines).  The selected time is after substantial emergence and just preceding the rapid rise and eruption of the emerged flux rope.  Notable features in both simulations are the sheath current at the top of the emerged flux rope, the quadrupolar magnetic structure in the $y=0$ plane following reconnection between the emerged field and background arcade, and the overall similarity (despite discrepancies in fine details) between the \DDsim{} and \GTsim{} simulations.  An animation of this figure \added{($53\unit{s}$; $t=0:200$)} is available in the online material, which shows the initial background arcade with the flux rope confined to the convection zone, followed by its rise into the corona and the eruption.  See text for detailed description of the dynamics.}
    \label{fig:3dOverview}
\end{figure*}

\begin{deluxetable*}{rlll}
\tablewidth{0pt}
\tablecaption{ \label{tab:units}}
\tablehead{
\colhead{Coeff.} & \colhead{S.I. Value} & \colhead{Note} }
\startdata
$L_N$ & $\rm 1.7\times10^{5}\unit{m}$  & Length \\
$\rho_N$ & $2.887\times10^{-4}\unit{kg}\unit{m}^{-3}$ & Density \\
$B_N$ & 0.13\unit{T} & Magnetic field \\
\hline 
$g_\text{sun}$ & 274.0 $\unit{m}\unit{s}^{-2}$  & $=B_N^2/\mu_0L_N\rho_N$ \\
$v_N$ & $6825\unit{m}\unit{s}^{-1}$ & $=B_N/\sqrt{\mu_0\rho_N}$\\
$t_N$ & $24.9\unit{s}$  & $=L_N/v_N$\\
$T_N$ & $7053.8\unit{K}$ & $=m_fm_pL_Ng_\text{sun}/k_B$ \\
$j_N$ & $0.609\unit{A}\unit{m}^{-1}$ & $=B_N/\mu_0L_N$ 
\enddata
\tablecomments{Normalizing coefficients and other parameters of the simulations.  Quantities below the line are derived from $L_N$, $\rho_N$, and $B_N$, and also make use of the proton mass $m_p\approx1.67\times10^{-27}\unit{kg}$, the mean particle atomic mass in the solar atmosphere $m_f=1.25$, and Boltzmann's constant $k_B\approx1.38\times{10}^{-23}\unit{J}\unit{K}^{-1}$.  Values accurate to double precision are used for $k_B$ and $m_p$ in the code.} 
\end{deluxetable*}

For our \GTsim{} simulation, we refer to the $\theta=\pi$ case from \citet{Leake:2022}, where the parameter $\theta$ will be described shortly.
The simulation models the emergence of a twisted flux tube from the model convection zone into the corona, and its subsequent eruption, using version \lare{} code \citep{Arber:2001}.
The full details of the equations solved, initial conditions, boundary conditions, and numerical details are given in that paper.
simulated emerging flux rope and observations of active regions on the Sun.
Here we summarize the results reported \citet{Leake:2022} that are required to compare the results of the \GTsim{} and DD simulations.

\autoref{fig:3dOverview} left provides a 3D perspective overview of the \GTsim{} simulation part way through the emergence process; an animation is available in the online material.
2D slices through three variables highlight the system's structure in terms of axial currents ($j_y$; blue-red scale), temperature ($T$; purple-yellow scale), and vertical magnetic field strength ($B_z$; gray scale).
Representative magnetic field lines are displayed as green lines.
The right panel displays the corresponding time from the DD simulation.
The vertical magnetic field in the $z=0$ plane is in the interior of the \GTsim{} simulation and at the lower boundary of the \DDsim{}.
The other two variables are shown in the $y=0$ plane, with $j_y$ shown for $x>0$ and $T$ shown for $x<0$.
Data in all cut planes are semi-transparent, allowing the field lines to be seen throughout the domain.

All data are reported in the normalized units used by \lare{} unless otherwise specified; see \autoref{tab:units} for their values in S.I. units.
The typical normalizing MHD variables of length ($L_N$), density ($\rho_N$), and magnetic field strength ($B_N$) are given at the top of the table; relevant derived coefficients are given in subsequent entries, along with their definitions.
In practice, the normalizing coefficient for density, 
$2.887\times{10}^{-4}\unit{kg}\unit{m}^{-3}$, was chosen so that gravitational acceleration of $g=1$ in normalized units corresponded to the solar value of $274\unit{m}\unit{s}^{-2}$.
Note that Equation (14) of \citet{Leake:2022}, defining the normalized temperature, is missing the factor of $m_f$ for the mean particle mass.
This does not affect either the simulation nor the analysis, is it only matters if certain physics modules, e.g., thermal conduction or partial ionization, are turned on during the simulation, which they were not.

The \GTsim{} simulation consists of a 3D Cartesian visco-resistive MHD simulation including gravity in the $\hat{z}$ direction.
The domain is initialized with a plane parallel, polytropic hydrostatic stratified atmosphere with $\gamma=5/3$ as the ratio of specific heats.
Stratification is defined in terms of a vertical temperature profile that includes an adiabatic convection zone ($z<0$), isothermal photosphere/chromosphere ($0<z<10$), exponential transition region ($10<z<20$), and isothermal corona ($20<z$), where distances are given in terms of the normalizing length $L_N=170\unit{km}$, approximately the photospheric pressure scale height of the Sun.
A stretched (non-uniform) numerical grid is used in both the horizontal and vertical directions such that, in physical units, the simulation spans $[-47,47]\times[-47,47]\times[-5,90]\unit{Mm}$\added{, resolved by $Nx\times Ny \times Nz = 512\times512\times768$ cells,} in $x$, $y$, and $z$, respectively.
The stretched grid is essentially uniform in the central portion of the domain and increases towards the side and top boundaries, and was intended to reduce the influence of these boundaries on the simulation interior (though the efficacy of this approach is suspect, depending on the simulation's evolution).

Superimposed on the stratified atmosphere is a magnetic field defined by two contributing substructures.
The first is the field due to a magnetic dipole moment positioned below the simulation's lower boundary (deeper in the convection zone), oriented such that it creates an arcade-like structure within the domain.
The arcade's polarity inversion line in the photospheric ($z=0$) plane, across which the vertical magnetic field passes through zero, is oriented along the $y$ axis, and the arcade loops are predominantly oriented in $x$-$z$ planes and point in the $+\hat{x}$ direction (see the outermost field lines in \autoref{fig:3dOverview} and compare to \citet{Leake:2022} Figure 3).
The magnetic field strength of the dipole field is spatially varying across the photosphere and has a typical value of $B=0.015$ ($20\unit{G}$ in physical units) in the strongest regions.

The second magnetic contribution is a cylindrical, twisted magnetic flux rope inserted into the convection zone.
The flux rope has both axial and azimuthal components generated by poloidal and toroidal current systems, and is characterized by the angle $\theta$ between the azimuthal component above the flux rope's axis and the background dipolar field (see \citet{Leake:2022} Figure 2).
We focus on the case where the flux rope axis is in the $-\hat{y}$ direction, and its azimuthal field above the axis points in the $-\hat{x}$ direction to make an angle $\theta=\pi$ with the dipolar field, such that the two flux systems have anti-parallel azimuthal components across the separatrix surface bounding the two magnetic domains. 
The guide field along the flux rope axis in $-\hat{y}$ has an initial peak magnitude of roughly double the peak azimuthal field inside the flux rope, a distance $r\approx2$ from the axis.
The relative strength of the axial and azimuthal components are such that the rope is marginally stable to the kink instability (see Equation 32 in \citet{Leake:2022}, and \citet{Linton:1996}).
This configuration allows substantial magnetic reconnection between the dipole and flux rope flux systems.

The flux rope has a nonzero Lorentz force, which is initially balanced via a pressure gradient generated by adding perturbations to the background stratification of density and temperature.
The pressure perturbation, $p_1$, is chosen to cancel the Lorentz force ($\nabla p_1 = \vect{j}\times\vect{B}$), and is translationally invariant along the tube axis. 
A variable entropy is specified along the tube axis so that the middle portion, near $y=0$, is buoyant, i.e., at that location, the tube has minimum in density but temperature equal to the background, so it is no longer in hydrostatic equilibrium. 
At the ends of the tube, the converse is true (minimum in temperature, density unchanged from background).
This choice causes only the middle of the rope to rise and form an $\Omega$-shaped flux tube that eventually breaks through the photosphere and expands into the corona.

\begin{figure*}
    \centering
        \includegraphics[width=\linewidth]{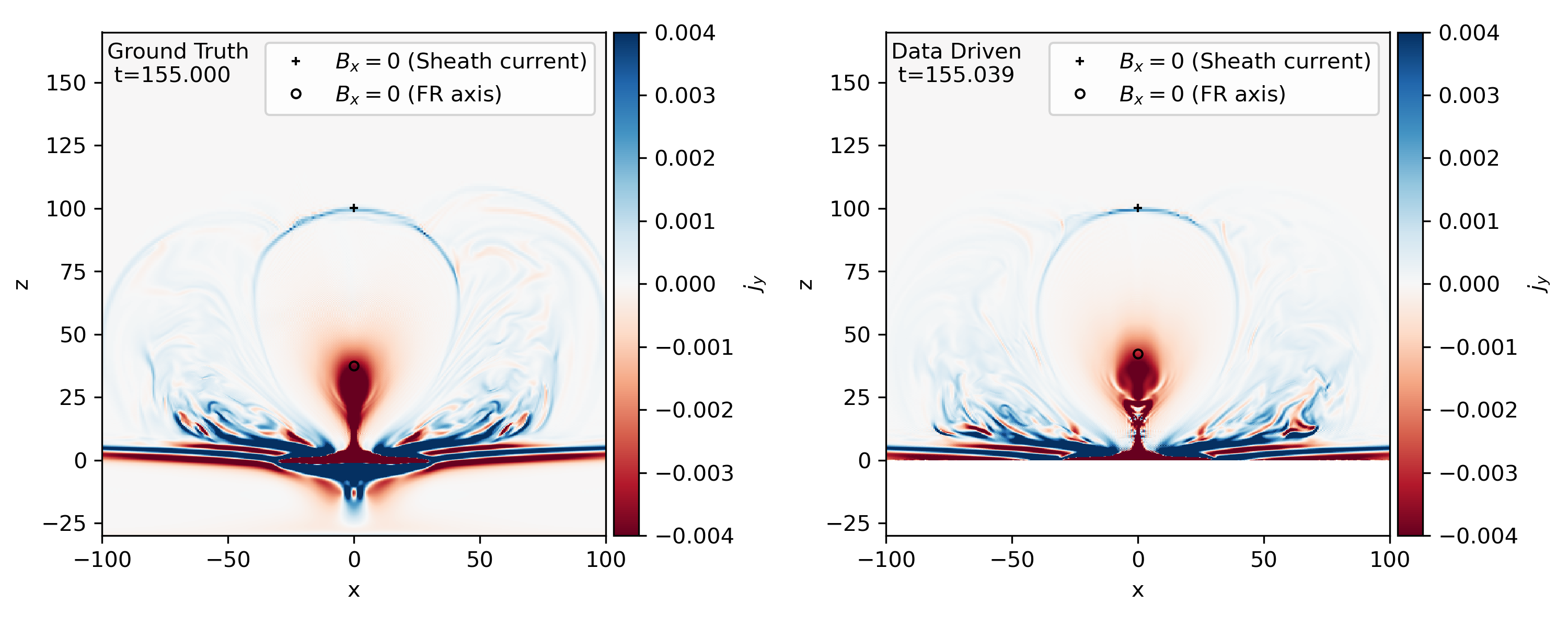}
    \caption{Axial current $j_y$ in the $y=0$ plane for a zoomed-in view of the emerging flux rope, in the \GTsim{} simulation (left) and \DDsim{} simulation (right).  The blank portion at the bottom of the \DDsim{} simulation lies outside of the \DDsim{} numerical domain (boundary at $z=0$).  The animation of this figure \added{($55\unit{s}$; $t=0:204.75$)} the shows the gradual rising and strengthening of the current sheets during emergence and rapid rise during the eruption.  The static figure corresponds to time $t=155$, shortly before the eruption in either simulation.  The cross and circle mark the locations of the sheath current and O-point along the $x=0$ line, respectively.}
    \label{fig:jyyeq0}
\end{figure*}

The snapshot of the \GTsim{} and DD simulations shown in \autoref{fig:3dOverview} is at time $t=150$, after substantial emergence, but prior to the eruption.
As \citet{Leake:2022} describe, during the early, gradual rise phase (up to about $t=120$), the rising and expanding twisted flux rope pushes upward through the photosphere and into the corona, displacing the background dipolar field.
A rising current sheet, associated with a topological separatrix surface, marks the interface between the two distinct flux systems.
This \emph{sheath} current appears as the enhanced positive (blue) axial current in the upper portion of \autoref{fig:3dOverview}.
In some circumstances the sheath current can be identified with a \emph{breakout current sheet} as described by the breakout model of \citet{Antiochos:1999}, for instance, during the eruption towards the end of the \GTsim{} simulation.
Note, however, that a breakout current sheet is the result of a potential arcade that is energized by photospheric shear, whereas our sheath current is present in the initial condition in the convection zone, and emerges through the photosphere into the corona.
Shear motions are not necessarily associated with the existence of a sheath current.
The processes involved in reconnection across the sheath current shift the separatrix surface upwards (a \emph{distinct} mechanism from the emergence process) while transferring flux to a set of side-lobes, eventually creating a roughly quadrupolar arrangement of flux in $x$-$z$ plane. 
This can be seen in field lines in the Figure and its animation.

The axis of the flux rope, as traced by field lines originating in the convection zone, never fully emerges into the corona.
Instead, a sheared arcade forms above the photosphere, and as it rises, a vertically oriented current sheet forms below it.
This is designated the \emph{flare current sheet} by \citet{Leake:2022}, and is obscured by the traced field lines in \autoref{fig:3dOverview}, but is visible as the vertical negative (red) $j_y$ region along $x=0$ in \autoref{fig:jyyeq0}.
Reconnection across the flare current sheet is associated with the reconfiguration of the field of the sheared arcade into a newly formed flux rope above the photosphere, which is distinct from the original flux rope.
The axis of the original flux rope remains confined to the convection zone and near-photospheric layer throughout the \GTsim{} simulation.
The ``axis'' that we will refer to in the remainder of the paper is the new axis that forms above the photosphere due to the combined effects of flux emergence, expansion, and reconnection across both the sheath and flare current sheets.

Many of the features just discussed can be easily identified in the axial currents ($j_y$) in the $y=0$ plane, shown in \autoref{fig:jyyeq0}, where the left panel is for the \GTsim{} simulation and the right panel is for the \DDsim{} simulation.
The heights of the sheath current and (new) flux rope axis are identified by zero-crossings of $B_x$ along the $x=y=0$ line and are marked by a $+$ and $\circ$, respectively.
The axis location has an O-point topology in the perpendicular ($x$-$z$) plane, i.e., the in-plane field circles around the O-point.
In 3D, however, a strong guide field in the $-y$ direction is present throughout the flux rope, even as the rope's axis writhes and rotates out of the $y-z$ plane as the system evolves.

Beginning around $t=150$, the sheath current, new flux rope axis (the O-point), and upper edge of the flare current sheet all rapidly rise higher into the corona, and there is continual reconnection across both the sheath and flare current sheets.
These dynamics are accompanied by rapid upward magnetic and mass fluxes.  
This event is the eruption, and its dynamics are distinct from the gradual rise phase that preceded it.
Indeed, the suite of simulations studied by \citet{Leake:2022} demonstrated that the onset of the eruption strongly depends on the relative orientation of the flux rope and dipolar field  (i.e., $\theta$), whereas the gradual rise phase is similar for all simulations, regardless of $\theta$. 

Further details and discussion of the $\theta=\pi$ and related simulations are presented in \citet{Leake:2022}; see especially their Section 3.1. 
The above summary is enough for our present purpose of comparing a DD simulation to the \GTsim{} simulation.
The animations of both \autoref{fig:3dOverview} and \autoref{fig:jyyeq0} already show that the dynamics in the DD simulation closely follow those of the \GTsim{} simulation, thus qualitatively validating our method.
Next we provide more quantitative comparisons.

\section{\GTsim{} Output and DD input: Extracted synthetic observations}\label{sec:syntheticBC}
The validation exercise we present here measures how well our method can handle data that is coarsely sampled in time. 
We have chosen a driving cadence similar to existing synoptic solar data, for example, from the Solar Dynamics Observatory \citep[SDO;][]{Pesnell:2012}, whose Helioseismic and Magnetic Imager \citep[HMI;][]{Scherrer:2012} provides estimates of the vector magnetic field across the Earth-facing surface of the Sun at a cadence of 12 minutes.
The emergence of a typical active region can last between 50-100 hours \citep[Ch.~3 of][]{Harvey:1993, Norton:2017}, which means HMI will typically provide 250-500 measurements over the entire emergence process.
Large scale eruptions tend to occur concurrent with or shortly after periods of rapid emergence.
Because the flux rope in the \GTsim{} simulation is nearly fully emerged by $t=100$ (see \autoref{fig:magflux}, discussed below), this makes the relative temporal scaling, in terms of the emergence dynamics, easy to compare between the simulations and observations.

Given the above discussion, to mimic actual observations, all MHD variables from the full numerical domain of the \GTsim{} simulation were saved at an approximate cadence of $\Delta t = 0.25$ from the initial condition up to $t=220$ (after the eruption).
These serve as the ``synthetic observations'' in the present test, both as input to our data driving algorithm, (using the photospheric layer only), and as output for comparison (using the full volume above the driven boundary layer).
See \autoref{sec:appendix} for details of variables in the extracted layer and the driving layer derived from them.
For the present study, we do not apply any noise or bias to these data---e.g., by radiative forward modeling of atomic lines and subsequent (re)inference of the MHD variables from those---but instead focus only on a realistically coarse temporal sampling of the system's dynamics.
See \citetalias{Tarr:2024} for the performance of our method under conditions of missing data and for a wide range of temporal sampling.

The base MHD code for both simulations, \lare{}, uses an adaptive time step ($\delta t$) calculated at each time to satisfy the dynamically evolving Courant-Friedrics-Lewy (CFL) condition \citep{Courant:1928}. 
This limits the time step to a fraction of the pixel crossing time for the fast wavespeed in the system; fast flows and shocks tend to reduce the time step.
A snapshot is whenever the simulation time passes each output $\Delta t$ period, so the variable time step leads to slight variations in the output cadence of the \GTsim{} simulation, and, hence, the input cadence to the \DDsim{} simulation. 
These variations are not important.

More important from an algorithmic standpoint is that the number of simulation steps between each \GTsim{} output time vary considerably over the course of the simulation, from 45 at the beginning to over 1300 during the eruption, with 1000 being a typical number after substantial emergence, due to the increased coronal field strength and corresponding increase in the Alfv\'en speed.
Therefore, updated boundary information is only provided to our data driving method at a cadence that is several orders of magnitude slower than the inherent, CFL-limited time step.
The method must successfully drive the system between these input times in a physically self-consistent manner.

Synthetic observations are created from the \GTsim{} simulation outputs by extracting all MHD variables at the single height of $z=0$, corresponding to the model photosphere.  
Appropriate averaging of each variable from the staggered grid used by \lare{} to a cell-centered grid was made during the extraction, so that all input variables provided to our data driving algorithm are spatially collocated.

\section{Data driven simulation} \label{sec:DDsim}
The \DDsim{} simulation is performed following the methodology described in \citetalias{Tarr:2024}.
\added{Its initial condition reproduces exactly the \GTsim{} simulation from the photosphere upwards, and therefore spans  $[-47,47]\times[-47,47]\times[0,90]\unit{Mm}$, resolved by $Nx\times Ny \times Nz = 512\times512\times648$ cells, in $x$, $y$, and $z$, respectively.  
Specifically, the $iz = 0$ layer of \DDsim{} (i.e., the layer of ghost cells immediately below the lower boundary of the standard numerical domain) is identified with the $iz=120$ layer of the \GTsim{} simulation.}
\lare{}'s ghost cells at the driven boundary are updated using our purpose-built characteristic MHD code, \charc{}. 
Bidirectional coupling keeps \lare{} and \charc{} in sync.
\charc{} solves an independent minimization problem at each boundary cell, and at each time, between a target MHD state and the set of accessible MHD states according to the characteristic description of MHD.
At each time step, the next ``target'' MHD state in the boundary layer is calculated by interpolating between the current values of primitive variables and those of the next driving input time (i.e., the data extracted from the \GTsim{} simulation).
We currently use a linear interpolation, but this could easily be replaced with a more advanced scheme, for instance, a higher order interpolation, or one resulting from a physics informed machine learning approach.

Once a target state for a given CFL step is defined, the MHD characteristics are used to find the update to each primitive variable in each boundary cell that most closely matches the target state, but restricted to those updates allowed by MHD. 
The updates are required to be consistent with (low-frequency) Maxwell's equations and Newton's laws, and setting up the problem in terms of the characteristics enforces that consistency.
Effectively, the characteristics are used to project the target state onto the space of accessible MHD states.
If the target state is in the space of allowed updates to the primitive variables (for instance, because a linear interpolation between the current state and the next driving state is a perfect approximation) then the method sets the boundary condition such that a standard numerical update of the boundary cells reproduces the target state.
If the target state is outside the space of allowed updates (for instance, because a linear interpolation is a bad approximation), then the method sets the boundary condition such that a standard numerical update in the boundary cells produces a state that is \emph{maximally close} to the target state by some minimization scheme.
In the present work, we again use a Singular Value Decomposition method to solve the minimization problem, as described in \citetalias{Tarr:2024}.
As with the linear interpolation of driving data, the optimization algorithm can easily be swapped out for a better method, but we do not explore that option here.
\autoref{sec:appendix} presents a comparison between the variables extracted from the \GTsim{} simulation and the boundary conditions derived from them by the \charc{} code.

Comparison of the right and left panels of the animation corresponding to \autoref{fig:3dOverview} shows that the dynamics of the \DDsim{} simulation closely follow those of the \GTsim{} simulation, as outlined in \autoref{sec:GTsim}, so we do not repeat that description here.
By itself, this simple comparison qualitatively demonstrates the validity of the method.
The main difference is in the timing of the eruption, with the flux rope in the \DDsim{} simulation erupting slightly before the \GTsim{} case, while minor differences can be found throughout.
The next section provides quantitative comparisons.

\section{Results} \label{sec:results}
\subsection{Unsigned magnetic flux}\label{sec:unsignedflux}
\begin{figure}
    \centering
    \includegraphics[width=1.0\linewidth]{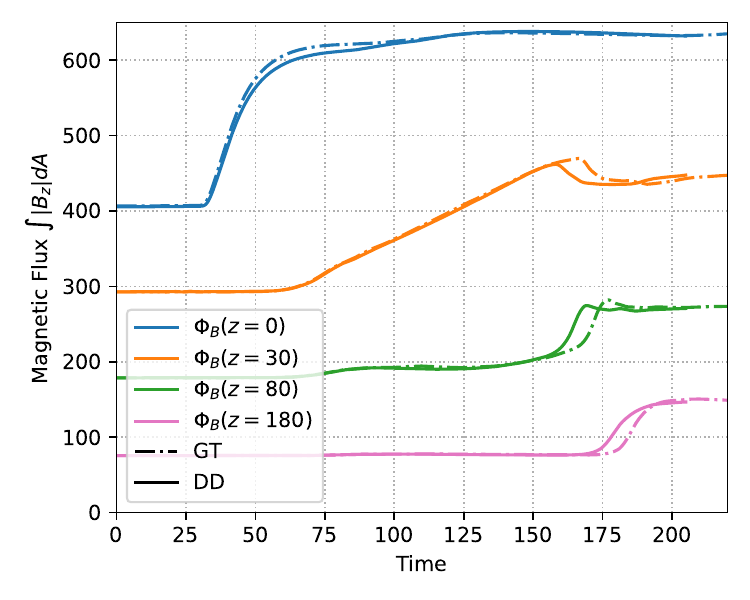}
    \caption{Unsigned magnetic flux integrated over constant-$z$ planes in the \GTsim{} (dash-dot) and \DDsim{} (solid) simulations, as a function of time.  The blue, orange, green, and pink lines correspond to the $z=0,30,80,180$ planes, respectively, and are sequential from top to bottom, as the flux decreases with height.  The two curves for each height mostly overlap.  The legend obscures only effectively constant portions of the $z=80$ and $180$ curves.}
    \label{fig:magflux}
\end{figure}

\autoref{fig:magflux} shows the unsigned magnetic flux through four constant $z$ planes as a function of time, where the blue, orange, green, and pink lines correspond to the $z=0,\ 30,\ 80,\ $ and $180$ planes, respectively. 
The typical decrease of flux with height is easily seen by the sequential ordering of the curves from top to bottom, with the values near the beginning of the time series representing the flux of the background arcade.
The flux through each layer increases as the emerging rope reaches that layer.
Solid lines correspond to the DD simulation, and dash-dotted to the \GTsim{} simulation.
The two curves mostly overlap in a given layer for much of the time series.
The initial sharp increase in the $z=0$ (blue) line around time $t=30$ marks the beginning of emergence through the photospheric plane, which is the lower, driven boundary of the \DDsim{} simulation and part of the interior of the \GTsim{} simulation.
The slower, linear rise in flux in the $z=30$ plane is caused by the expansion of the emerged rope into the corona.
The eruption is visible as the sudden drop in flux within the $z=30$ plane (orange, $t\approx 160$), followed by the subsequent rapid arrivals of flux in the $z=80$ and $z=180$ planes (green and pink, respectively).
The eruption in the \DDsim{} simulation (solid) precedes that in the \GTsim{} simulation (dash-dot). 

\subsection{Mass flux}\label{sec:mass}
\begin{figure*}
    \centering
    \includegraphics[width=\linewidth]{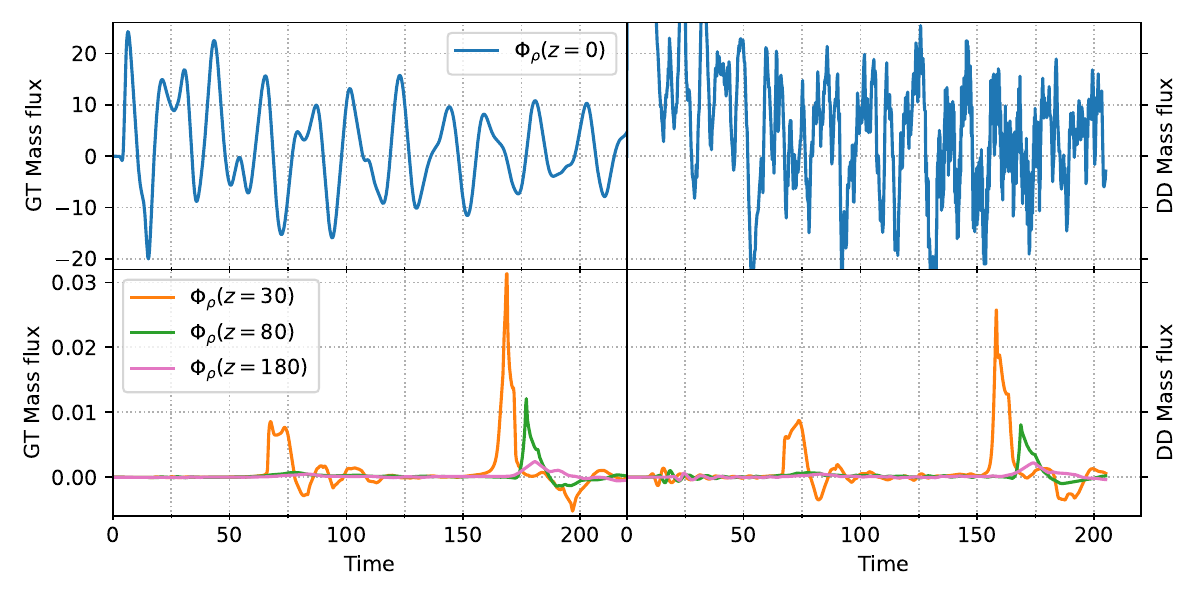}
    \caption{Net mass flux integrated over constant-$z$ planes as a function of time, with the \GTsim{} values at left and the \DDsim{} results at right.  The upper panels show the mass flux through $z=0$ plane (blue), and the lower plots show flux through the $z=30, 80,\ 180$ planes in orange, green, and pink, respectively.  The scales are the same for both simulations, but differ by 3 orders of magnitude for the photospheric (upper) versus coronal (lower) calculations. Note that the \GTsim{} data in this uses a solid line, unlike other the figures.    }
    \label{fig:massflux}
\end{figure*}

\autoref{fig:massflux} shows the mass flux through the same four constant-$z$ layers ($z=0, 30, 80, 180$) over time, using the same color scheme.
The upper panels record the flux through the photosphere, and the lower panels show flux through the successively higher layers of the corona; note that the scale of the lower plot is 3 orders of magnitude less than the upper plot.
Comparing the left and right panels of the lower plot clearly demonstrates that the \DDsim{} simulation captures complex aspects of the mass flow throughout the coronal portion of the simulation.
In particular, it reconstructs features of the wake of the eruption.
At the lowest coronal height ($z=30$; orange), the eruption takes off around $t=165$ (or slightly earlier, $\approx160$, in the \DDsim{} simulation), with a large net upward flow of material.
This is followed by a much smaller amplitude downflow in the wake of the eruption ($t=185$) and a subsequent rebound ($t=195$).
The timing is earlier in the \DDsim{} simulation, as the eruption is triggered earlier, but the amplitude of each pulse is within $20\%$ of the corresponding \GTsim{} value.

While the mass flux compares favorably in coronal regions of the simulations, the \DDsim{} simulation contains high frequency variations in the driving layer that are not present in \GTsim{} simulation (upper panel in the figure, blue lines; see also \autoref{sec:appendix}).
However, a few additional metrics suffice to demonstrate that, despite the high frequency differences, the mass flux in the driving layer does agree quite well between the two simulations.
For the rest of this subsection, we exclude times before $t=25$ because the \DDsim{} simulation contains initial transient flows triggered by slight imbalances when matching the cell-centered numerical formulation of \charc{} to the staggered grid of \lare{}. 
The system stabilizes after the initial transients damp away.

First, between $t=25$ and the $t=204.75$ (the last output time of the \DDsim{} simulation), the mean mass flux is $+1.5$ mass units per unit time for \GTsim{} and $+4.5$ for \DDsim{}, compared to a total mass in the $z=0$ layer that is of order $10^6$.
Second, the peak of the cross correlation between the two time series in the top panels of \autoref{fig:massflux} occurs with a time-lag of $\Delta t = 0.25$ (a single output time).
Despite the high-frequency differences, the driven mass flux does track the slow oscillations present in the original mass flux.
Finally, Fourier transforms of both time series reveal the same three dominant periodicities of $\tau = 20, 11.25,$ and $7.82$ in normalized time.

The differences between the two mass fluxes at the driving layer are likely caused by peculiarities of our optimization technique and a degeneracy in the solution space for density, entropy, and pressure.
This causes some non-smooth pixel-to-pixel variations in the thermodynamic variables, which in turn generate small shocks that cause the high-frequency variations seen in the timeseries.
We discuss this further, as well as possible solutions, in \autoref{sec:discussion} and \autoref{sec:appendix}.

\subsection{Poynting flux}\label{sec:poynting}
We calculate the net Poynting flux, as a function of time, through the same four constant-$z$ planes,
\begin{align}
    P_z(z;t) & = \int dx dy \vect{S}(z)\cdot\hat{z} = \int dx dy( \vect{v}\times\vect{B})\times\vect{B}\cdot\hat{z}\\
    & = \int dx dy (\vect{v}_\perp\cdot\vect{B}_\perp B_z - B_\perp^2 v_z)\label{eq:poyntingflux},
\end{align}
where $\vect{S}=\vect{E}\times\vect{B}=(\vect{v}\times\vect{B} )\times\vect{B}$ is the Poynting vector using the electric field and $\perp$ refers to vectors in the horizontal ($x-y$) direction.
The computation is performed in the form given in \autoref{eq:poyntingflux}, using values of the magnetic field interpolated to cell vertices, which are the natural locations of the velocity variables in \lare{}.

\begin{figure}
    \centering
    \includegraphics[width=1.0\linewidth]{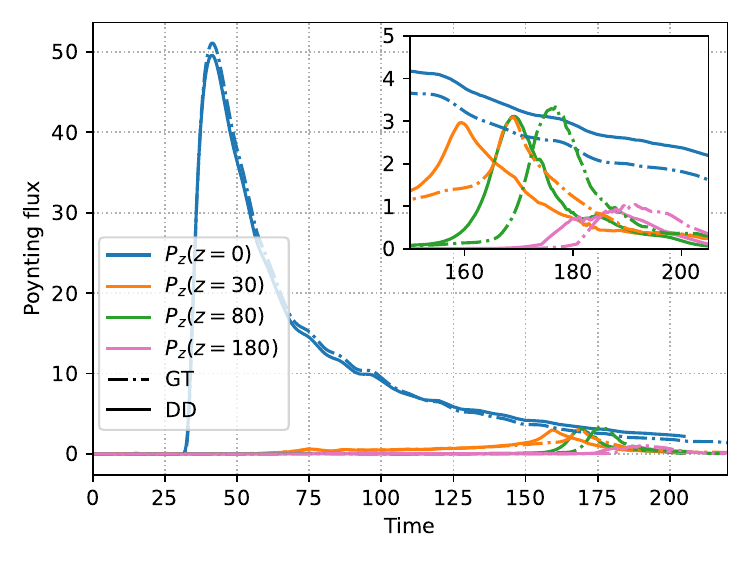}
    \caption{Net Poynting flux calculated through four constant-$z$ layers, $z=0,30,80,180$ for blue, orange, green, and pink curves, respectively, for the \GTsim{} (dash-dot) and \DDsim{} (solid) simulations, as a function of time.  The blue curve ($z=0$) shows the emergence of the flux rope through the photosphere, while pulses in the orange, green, and pink curves starting around $t=160$ show the successive transport of magnetic energy into higher portions of the atmosphere during the eruption.}
    \label{fig:poynting}
\end{figure}

The vertical Poynting flux quantifies the net transport of electromagnetic energy by the plasma in the $\hat{z}$ direction, and is therefore expected to show strong signals from both the emergence and the eruption of the flux rope.
\autoref{fig:poynting} demonstrates that this is indeed the case. 
The line and color schemes are the same as in \autoref{fig:magflux}.
The blue curve ($z=0$) primarily represents the transfer of magnetic energy into the domain above the photosphere as the flux rope emerges.
The peak amplitude during emergence is slightly suppressed in the \DDsim{} simulation (solid), but the timing is essentially the same, both in the photosphere (blue; $t=45$) and low corona (orange; $t=75$).
However, we once again see that the eruption begins slightly early in the \DDsim{} simulation, at $t\approx160$ for $z=30$, as opposed to $t\approx170$ in the \GTsim{} simulation.
The relative timing of the eruption at subsequent heights (green and pink curves), and the amplitudes of the vertical Poynting flux at all heights, are basically the same between the two simulations, which is evidence that the \emph{dynamics} of the eruption are well reproduced by the \DDsim{} simulation. 

\subsection{Topological proxies of erupting flux system}\label{sec:topology}
Topological comparisons between 3D MHD simulations can become complicated.
We have therefore chosen to compare a few simple metrics that are closely related to topological features of the system: (i) the axial current $j_y$ in the $y=0$ plane, and (ii and iii) the vertical location of the sheath current and emerged flux rope axis, respectively, along that the $x=y=0$ line.

\begin{figure*}
    \centering
    \includegraphics[width=\linewidth]{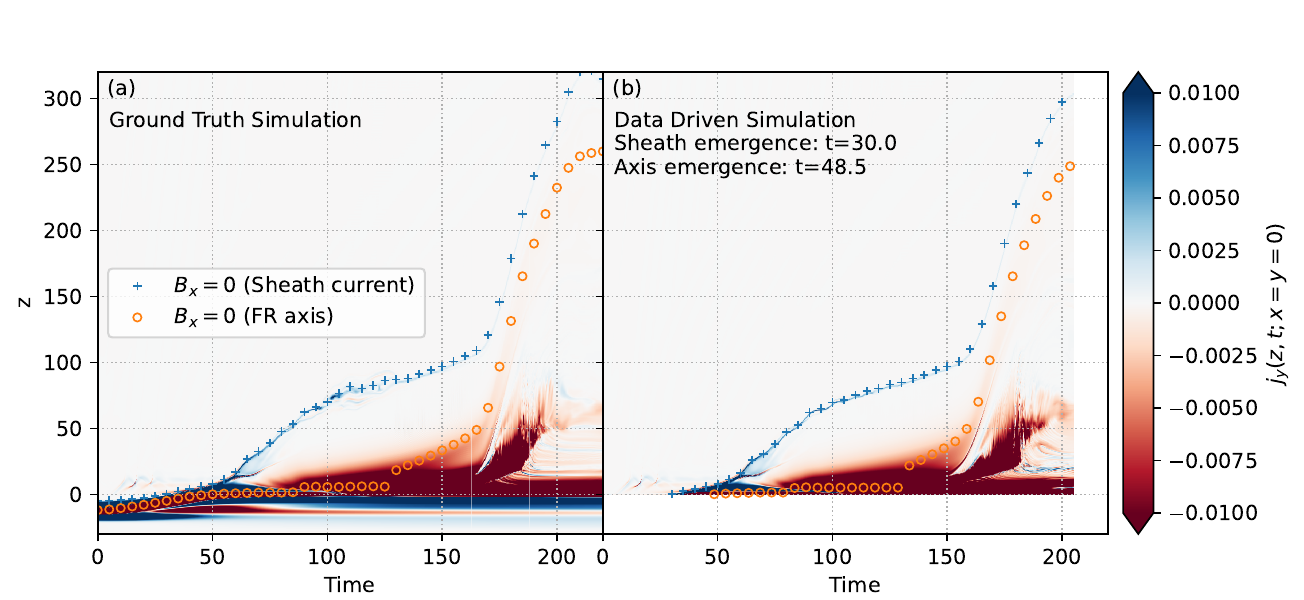}
    \caption{Space-time diagram of axial current density $j_y$ extracted along the $x=y=0$ line (red-blue contour plot), for the \GTsim{} (left) and \DDsim{} (right) simulations.  Blue pluses and orange circles mark the locations of the sheath current and flux rope axis (FR axis), respectively, identified by zero crossings of $B_x$ along the same line.  Symbols are plotted with a temporal spacing of t=5 (beginning with their first appearance in the domain, for the \DDsim{} case) to avoid obscuring the current density.}
    \label{fig:jyxt}
\end{figure*}

\autoref{fig:jyyeq0} and its animation provide useful context, showing $j_y$ in the $y=0$ plane and the locations of the sheath current and flux rope axis.
That figure also includes the location of the sheath current ($+$) and flux rope axis, or O-point ($\circ$), identified as zero-crossings of $B_x$ along the $x=y=0$ line.
As stated earlier, both features are present in the convection zone in the \GTsim{} simulation before emerging through the photosphere, but are not present in the initial condition of the \DDsim{} simulation.

\autoref{fig:jyxt} is a height-time plot of $j_y$ along the $x=y=0$ line in the \GTsim{} simulation (left) and \DDsim{} (right).
The axial current density shows many fine evolutionary features common to both simulations.
Overplotted are the heights of the sheath current (blue `$+$') and flux rope axis (orange $\circ$), which is also approximately the upper edge of the flare current sheet.
Both of the latter two features were identified using locations where $B_x=0$ along that line.
Note that neither the sheath nor the axial zero crossing exists in the \DDsim{} domain until $t\approx30$ because the original buoyant flux tube is initialized within the \GTsim{} simulation's convection zone.
The symbols are plotted as each feature enters the driven domain at the times indicated in the right hand panel ($t=30$ for the sheath current and $t=48.5$ for the axis).
The sheath current is thin enough and matches the location of the upper $B_x$ zero crossing well enough that the symbols for the latter can obscure the former; we therefore plot symbols at a spacing of $dt=5$.
The flux rope axis has a less localized current distribution, though in that case the $B_x$ zero crossing strongly tracks the upper edge of the axial current system.

Starting with the eruption ($t\gtrsim 160$), numerous additional zero crossings form as the flare current sheet thins and breaks up.
This is seen in both simulations, but we have not included additional symbols as they would further obscure the plots of the current density.
Many fine details can be found that agree in both plots, showing that, while differences do exist, the gross features of the current systems are largely the same in both the \GTsim{} and \DDsim{} simulations.
The major difference, as noted before, is the timing of the eruption, which occurs earlier in the \DDsim{} simulation.

Finally, the jumps in the O-point (orange $\circ$) in \autoref{fig:jyxt} around $t=80$ and $t=130$ are due to sudden shifts in the locations of $B_x$ zero point crossings in time.
These jumps are real, and have to do with folds in $B_x$ that propagate upward or downward through the domain.
The folds can have several causes, either associated with the formation and/or bifurcation of current sheets, or as a result of small shocks propagating through the system.
These processes only affect the O-point region two times during the simulations, but are very common (many tens of examples) after the eruption begins at late times in the lower portion of both simulations, where they are associated with reconnection across the flare current sheet.
For the O-point, the precise timing and magnitude of the two jumps are slightly different in the \GTsim{} versus DD simulation, but overall quite similar.

\subsection{Energy}\label{sec:energy}
\begin{figure}
    \centering
    \includegraphics[width=\linewidth]{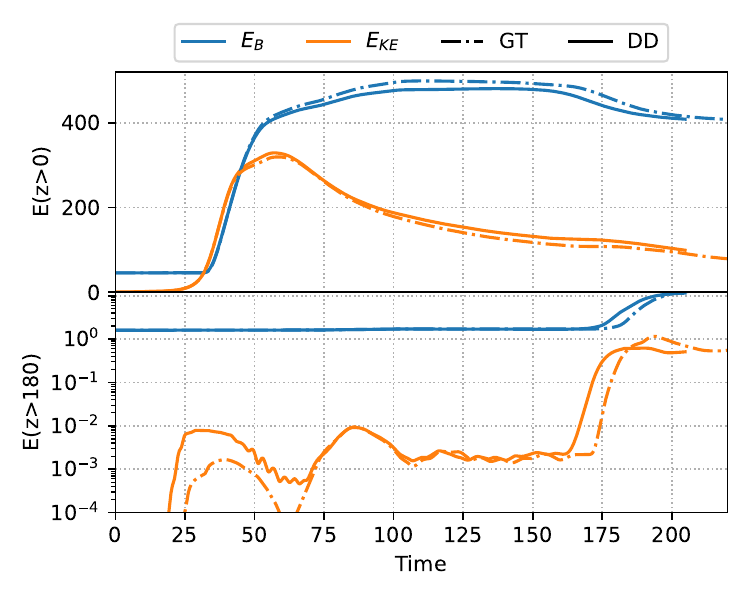}
    \caption{Total magnetic (blue) and kinetic (orange) energies in overlapping volume of the \GTsim{} (dash-dotted) and \DDsim{} (solid) simulations, as a function of time.  The upper panels show the calculation for the full volume above the driven boundary ($z>0)$ on a linear scale, while the lower boundaries show the values high in the corona ($z>180$) on a log scale to highlight the dynamics of the eruption.}
    \label{fig:energies}
\end{figure}

Finally, we calculate the total integrated magnetic ($E_B=\frac{1}{2}\int B^2 dx^3$) and kinetic ($E_\text{KE}=\frac{1}{2}\int\rho v^2 dx^3$) energies in the volumes above $z=0$ (upper panel) and $z=180$ (lower panel) in \autoref{fig:energies}, for the \GTsim{} (dash-dotted) and \DDsim{} (solid) simulations as a function of time.
The minor differences between the \GTsim{} and \DDsim{} simulations are mostly discernible high in the corona and early in the simulation ($t<100$) due to the log-scale of the lower panel.
Once again, the main difference is the timing of the eruption, which takes off earlier in the \DDsim{} simulation.
After emergence, the amplitudes of the energies agree to within a few percent between the simulations, except briefly in the upper corona due to the difference in onset of the eruption.
Even in that case, a simple temporal shift of one curve onto the other shows good agreement.

We defined the total energy injected during emergence as the difference of the magnetic and kinetic energies above $z=0$ (upper plot) between times $t=125$ and $t=0$, where $t=125$ corresponds to the time at which the integrated vertical magnetic field through the $z=0$ plane stabilizes prior to the eruption (see \autoref{fig:magflux}).
Between those times, the magnetic and kinetic energies of the \DDsim{} simulation increased by $434$ and $154$, respectively, as compared to $453$ and $141$ for the \GTsim{} simulation.
The respective sums of magnetic and kinetic energies, of $588$ (\DDsim{}) and $593$ (\GTsim{}), match to within $1\%$.

Similarly, we defined the energy of the eruption as the difference between the integrated energy densities above $z=180$ (lower plot) just after the eruption ($t=203$) and before ($t=160$) the eruption.
These times are chosen to match the (roughly) steady states of both simulations outside of the eruption time.
The eruption is dominated by changes in magnetic energy, with 10.5 for the \GTsim{} versus $9.75$ for \DDsim{}, compared to changes in the kinetic energies of just $0.72$ and $0.5$, respectively.
The respective sums agree within about $10\%$ for the two simulations, and in both cases represent about $2\%$ of the total energy transferred through the photosphere during emergence.

\section{Discussion and Conclusions} \label{sec:discussion}
Our goal in this study is straightforward: to establish that our data driving method, as fully described and validated on a toy problem in \citetalias{Tarr:2024}, is capable of performing well on the far more complicated and nuanced situations expected for solar use cases.
Our test case is the previously studied, ab initio flux emergence simulation of \citet{Leake:2022}, which includes gravity, a stratified atmosphere, a buoyant flux rope, and an overlying magnetic arcade. 
That simulation produced a variety of complicated, interesting dynamics that included the formation of a new flux rope inside the system and a rapid eruption late in the simulation.
We found that our \DDsim{} simulation is able to reproduce all of these features, across a variety of metrics, with high fidelity.

Any data driven method, when applied to solar observations, will typically require the use of \emph{sparsely sampled} data.
The data are likely sparse across many dimensions: spatially, temporally, or in terms of the sampled variables that constitute the MHD state vector.  
For example, the line-of-sight magnetic field is typically better constrained from spectropolarimetric observations than, say, the density or horizontal velocities.
In the present case, we have focused only on temporal sparseness that is comparable to current synoptic solar observations.
If one considers the emergence time of the flux rope in the \GTsim{} simulation, most of the flux has emerged through the simulated photosphere by $t=50$.
A typical active region on the Sun emerges most of its flux in about two days, suggesting that, dynamically, the unit time in our simulation is roughly equivalent to one hour of solar evolution.
The cadence of $\Delta t =0.25$ for extracting photospheric data from the \GTsim{} simulation and providing it to the \DDsim{} simulation data then corresponds to about 15 minutes of solar time.
This is slightly longer than the 12 minute cadence of the standard SDO/HMI vector magnetic field data series and indicates that our method could be applied to such data.

We did not consider numerous other types of sparsity in the driving data, nor did we consider the effects of missing data, noise, or bias, all of which are common features of solar observations.
Some of these additional issues were considered in \citetalias{Tarr:2024}, see especially their Sections 6 and 7.  
Further testing is required quantify errors associated each of those effects in data driven simulations, and how they compare to other sources of error and uncertainty.
For instance, the choice of initial condition is perhaps the most critical input to any simulation, and can result in vastly different system trajectories with comparable driving techniques.  
\citet{Chen:2023} and \citet{Wagner:2024} have each explored some of these issues, especially the variations in the initial condition, within their respective data driving methodologies.
Interestingly, \citet{Wagner:2024} compared a set of magnetofrictional models to zero-$\beta$ MHD models that are initialized with the same initial magnetic field and found that the MHD models are significantly more eruptive.
Contrarily, \citet{Barnes:2024} found that the initial condition had little effect on end state of a set of MHD relaxation experiments, and instead argued that the side and top boundary conditions were more important.
Clearly, carefully designed numerical experiments are necessary to disentangle these competing sources of error.

While the metrics in \autoref{sec:results} show that our method is able to reproduce both the gross features and many fine details of the \GTsim{} simulation, there are some notable differences.
First, there is the timing of the eruption, which occurs slightly earlier in the \DDsim{} simulation compared to the \GTsim{} simulation.
Tracking down the precise reason for this may be difficult, but we suspect that it is due to shorter simulation time steps in the \DDsim{} simulation and the resulting increase in numerical diffusion of the magnetic field. 
The shorter time steps, sometimes an order of magnitude less than the \GTsim{} time step, result from non-smooth solutions to the optimization problem in the driven boundary, particularly in the density, as alluded to at the end of \autoref{sec:mass} and detailed further in \autoref{sec:appendix}.
Indeed, the top left panels of \autoref{fig:boundary} show the overall good agreement between the average density structures in the two simulations, but with single pixel spikes present in the \DDsim{} simulation throughout portions of the domain.
We did not encounter this issue in the test problems described in \citetalias{Tarr:2024} or \citetalias{Kee:2025}.
Determining its cause will be the focus of future work.
The effect, however, is readily apparent: the single pixel spikes in the solution for the density (and associated spikes in the internal energy) generate many small shocks, whose short length scales require short time steps through the CFL stability condition.  
Another consequence is to produce the more complicated mass flux at the \DDsim{} lower boundary seen in \autoref{fig:massflux}, whereas the associated plot for the \GTsim{} simulation in the left hand panel is smooth.

There are several potential ways to fix this issue.
The first might be to lift a degeneracy in the optimization solution space between density, specific internal energy density, pressure, and entropy.
Right now the equations are solved in terms of density and specific internal energy.
Switching to a density and pressure representation, and then specifying a set entropy for all up-flowing material, may help resolve the issue.
Such an approach is similar to how inflows are treated in some magnetoconvection simulations \citep{Rempel:2014}.

Another possibility is to add a spatial regularization term to the optimization step.
That, however, is not as simple as it seems.
While we desire relatively smooth solutions in terms of the primitive variables, the optimization is carried out in terms of the characteristics, which are manifestly \emph{not} spatially smooth: infinite-gradient boundaries exist everywhere in the characteristic description of the system.
So, adding a regularization term likely requires some care.

Third, as mentioned in \citetalias{Tarr:2024}, we are currently using a singular value decomposition (SVD) approach to solve the optimization problem, which does add a slight level of inconsistency to our method (see discussion at the end of Section 5 in that paper).
SVD could be substituted out for a nonlinear optimization method.
Our tests of such an approach are not yet stable: eventually the system is driven to a state from which the optimization routine cannot find a converging solution.
More work is required to explore the nonlinear optimization options.

Finally, our method requires some estimate of the entire MHD state vector over a surface as a function of time, and the question of how best to derive those from existing data sources remains unanswered.
Our tests in \citetalias{Tarr:2024} showed that it is possible to achieve decent results for data driving when providing information of just a subset of the MHD variables, i.e., only the velocity and magnetic field.
Estimates for the other variables can be inferred using more extensive spectral diagnostics, for instance, in an approach analogous to the IRIS2 tool described by \citet{SainzDalda:2019} for ultraviolet observations.
The recent interest in physics informed neural networks (PINNs) may offer yet another method for deriving approximately self-consistent estimates of physical boundary conditions for the data driving problem, which could then be supplied to our method to make their evolution fully compatible with MHD.
The efficacy of the first step in such an approach, at least for the hydrodynamic case, is evident in the work of, e.g., \citet{Keller:2025}, who was able to use a PINN to recreate the 3D thermodynamic state of simulated hydrodynamic photospheric granulation using only a time series of synthetic continuum intensity images.

\added{Many other tests are necessary to further understand the robustness of our (and others') data driven codes, and we highlight a few here that we intend to address in the near future.
Cross-validation between MHD codes should be undertaken to determine if results are valid in the face of different numerical schemes with different forms of numerical diffusion.
So far, we have only compared a ground truth \lare{} to a driven \lare{} simulation. 
The same is true for most of the other data driving studies cited in Introduction, e.g., \citet{Chen:2023} compares ground truth to driven MURaM \citep{Vogler:2005, Cheung:2019} simulations.
We are currently testing the use of the driving data from the present simulation to drive an ARMS simulation \citep{Devore:2008} using the modular version of the \charc{} code mentioned in the Introduction, and plan to report on that effort in a future publication.
A complementary test would be to drive \lare{} with data extracted from, say, a MURaM simulation that includes photospheric convection with its ever-evolving mix of boundary inflows and outflows.
Our approach should be able to handle that case, and indeed, the present simulation already has a fully open lower boundary that contains a mix of inflows and outflows, as shown in the animation of \autoref{fig:boundary}.
But, this would be a yet more stringent test that includes additional complexities of real observations.}

\added{An open question is what physical resolution is required from photospheric observations to accurately reproduce coronal dynamics.
This is not a straightforward test to perform.
A typical approach for numerical experiments is to run the same experiment at different resolutions---spanning a few factors of two, say---and look for convergence.
\citet{Chen:2023} performed such a test and found that some dynamics and metrics seemed to match quite well between ground truth and driven simulations, while others did not.
This highlights the challenge of observationally driven simulations: the information content at the boundary is limited by the observations, not the simulations (these simulations are all far from reaching the limit of the world's largest super computers).
We speculate that what matters most is whether the small-scale features being removed by a coarser resolution (or perhaps just smoothing) have an outsized influence on either the global topology of the system or particularly important current structures.  
For instance, there are some theories that posit the importance of small-scale flux emergence at specific locations with specific orientations for triggering eruptions \citep{Kusano:2020}.
But, it is difficult to know which small scale features may be important and which are not, or if important sub-resolution features exist in current observations.
For the present case, the \GTsim{} simulation is converged, and the physical Courant condition introduced in \citet{Leake:2017} and discussed in \citetalias{Tarr:2024} section 7.1, i.e., the ratio the smallest feature's horizontal length scale to its horizontal velocity, is well satisfied.  We would therefore not expect reduced spatial resolution to greatly affect the general properties of dynamics, though this should be verified in future tests.}

\added{Perhaps the most important test from an observational perspective is to add realistic noise to the input data.
This is particularly true for the horizontal magnetic and velocities fields, which are much less well constrained than the line-of-sight components of either field.
This also has direct implications for the science requirements of upcoming observatories that could provide data for driven simulations in the future, notably, ngGONG (\url{https://nso.edu/telescopes/nggong/}).}

While there are clearly numerous avenues for improvement, the current work represents a major step forward in the development of data driven MHD simulations. 
As far as we are aware, this is the first fully data driven simulation to operate wholly within the full MHD framework that reproduces an entire active region evolution sequence from pre-emergence to eruption, and does not include any other ad hoc assumptions or fine-tuning of parameters.
That it does so with high fidelity across a variety of metrics, including mass, energy, and Poynting fluxes, the reproduction of important topological features, and the presence of an eruption, gives us high confidence that it can produce reliable results when applied to solar observations.

\begin{acknowledgments}
This work is supported by the National Solar Observatory, the Office of Naval Research, and the NASA HSR and LWS Programs.
N.D.K acknowledges support from the H-ISFM program ``Development of Novel Data-Driven Modeling Capabilities'' and the NASA Internal Research and Development (IRAD) program ``Developing the Machinery for in-situ Data-Driven Simulations.''
M.G.L. acknowledges support from the Office of Naval Research. 
P.W.S. and M.G.L. acknowledge support from NASA grant NNH21ZDA001N-LWS ``The Origin of the Photospheric Magnetic Field: Mapping Currents in the Chromosphere and Corona.''
M.G.L. acknowledges support from NASA grants NNH18ZDA001N-HSR ``Investigating Magnetic Flux Rope Emergence as the Source of Flaring Activity in Delta-Spot Active Regions,'' and NNH20ZDA001N-HSR ``Investigating the Influence of Coronal Magnetic Geometry on the Acceleration of the Solar Wind.''
P.W.S. and J.E.L. acknowledge support from the NASA Internal Science Funding Model (H-ISFM) program ``Magnetic Energy Buildup and Explosive Release in the Solar Atmosphere.''
This research has made use of NASA's Astrophysics Data System.
Resources supporting this work were provided by the NASA High-End Computing (HEC) Program through the NASA Advanced Supercomputing (NAS) Division at Ames Research Center.
\end{acknowledgments}

\begin{contribution}

L.A.T. was responsible for developing and running the data driven simulations, performing the initial analysis, and writing and submitting the manuscript.
N.D.K. was heavily involved in the development of the characteristic boundaries and their numerical implementation.
J.E.L. led the development and analysis of the original Ground Truth simulation.
P.W.L. and M.G.L. provided feedback on development of the problem, critical review and commenting on the manuscript, and much of the initial impetus to develop these data driven simulations.
All authors contributed to the analysis of the results, development of figures, and comments on and improvements to the manuscript.

\end{contribution}

\software{Simulations performed using the \lare{} \citep{Arber:2001} and \charc{} \citep{Tarr:2024} MHD codes. Analysis primarily carried out in Numpy \citep{Harris:2020} in python3.  Figures created with Matplotlib \citep{Hunter:2007}, and Tecplot (\autoref{fig:3dOverview}).}

\appendix
\section{Lower boundary details}\label{sec:appendix}
\begin{figure}
    \centering
    \includegraphics[width=\linewidth]{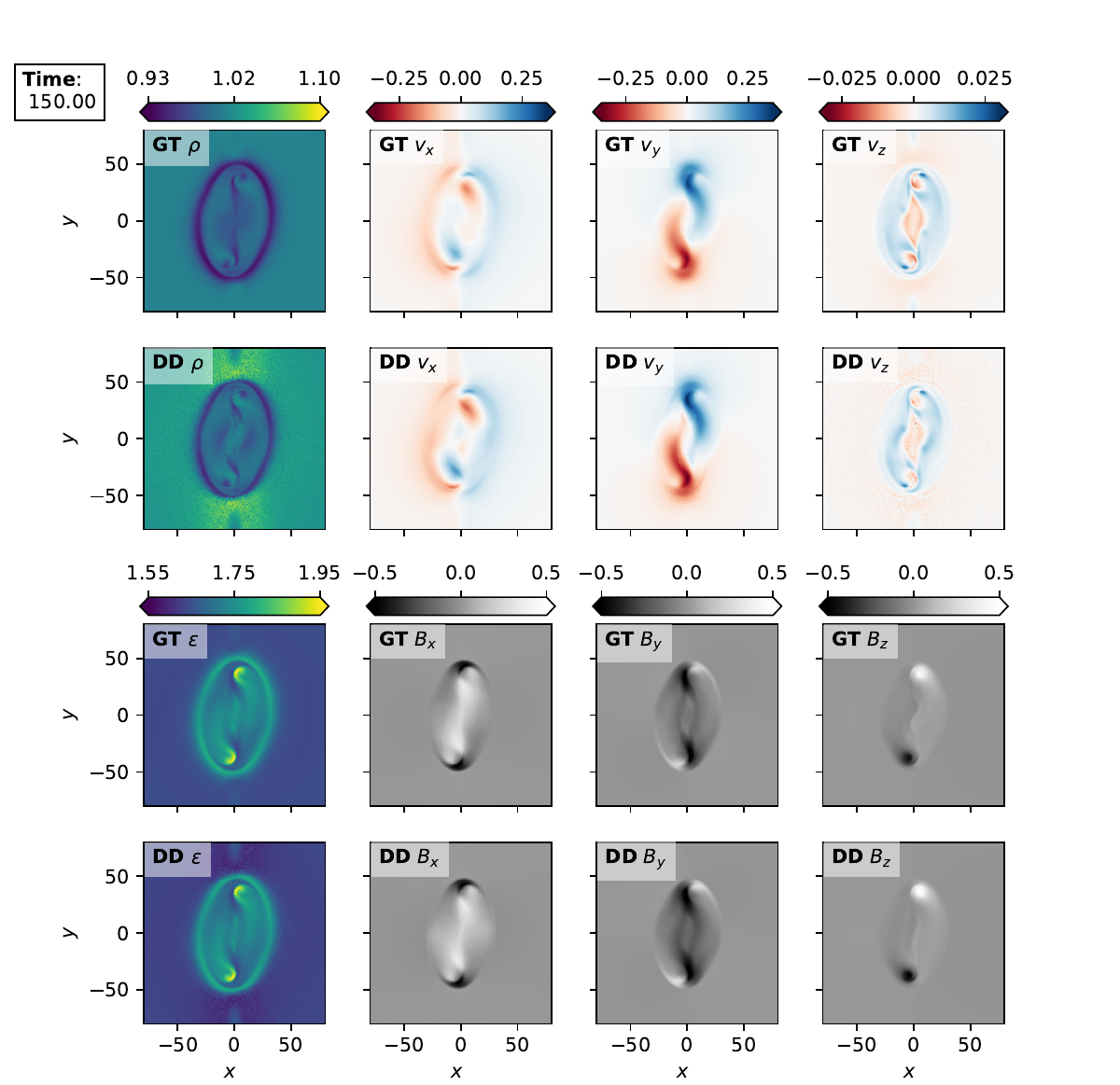}
    \caption{Primitive variables in the central portion of the $z=0$ layer at time $t = 150$.  Rows 1 and 3 show data extracted from the \GTsim{} simulation and used as inputs to the \charc{} data driving code; rows 2 and 4 show the outputs of the \charc{} code, which are the boundary conditions for the \DDsim{} simulation.  The first column shows the density (upper) and internal energy density (lower), and the next three columns show the three components of the velocity field (upper), and magnetic field (lower).  The output of the data driving algorithm generally matches the input (target) ground truth data to a high degree, though variations can be found throughout, most notable for the density.  An animation of this figure ($55\unit{s}$; $t=0:204.75$) is available in the online material, which initially shows the magnetic arcade, and then the bipolar emergence of the rising flux rope.  Note that the scale for $B_z$ increases during the animation, as the emerged field is substantially stronger than the background arcade.}
    \label{fig:boundary}
\end{figure}

\autoref{fig:boundary} details the driven boundary in the $z=0$ layer. 
It is cropped horizontally to the central region of the simulation that encompasses most of the action.
Rows 1 and 3 show the ground truth data for all primitive variables extracted from the \GTsim{} simulation.
These define the target data that the data driving code attempts to reach at each time step.
Rows 2 and 4 show the output of \charc{} code, i.e., the closest MHD state to the target state, given the current state of the boundary (including, implicitly, its time history) and the updates allowed by incoming characteristics.
Column 1 shows the density and internal energy density in the upper and lower panels, respectively, and columns 2, 3, and 4 show $(v_x, v_y, v_z)$ and $(B_x, B_y, B_z)$ in the upper and lower panels, respectively. 
Most of the variables show very little difference between input and output, though minor differences can be found throughout.
As discussed in the main text, the largest differences are found in the density and, to a lesser extent, in the internal energy density.
These take the form of broad regions with mottled, sharp features, on the scale of individual pixels, that we suspect are due to near-degeneracies in the solution space of the minimization algorithm.
These non-smooth pixel-to-pixel variations in density generate numerous small shocks throughout the driving layer and seem to have many carry-on effects, most notably short CFL time steps, high frequency mass flux, and the non-smooth vertical flows seen in the $\mathbf{DD}\ v_z$ panel (second row, rightmost column).

\bibliography{2025_DDBC_p3}{}
\bibliographystyle{aasjournalv7}

\end{document}

%% file: macros.tex
\DeclareRobustCommand{\okina}{%
  \raisebox{\dimexpr\fontcharht\font`A-\height}{%
    \scalebox{0.8}{`}%
  }%
}
\newunicodechar{ʻ}{\okina}

\defcitealias{Tarr:2024}{Paper I}
\defcitealias{Kee:2025}{Paper II}

\newcommand{\fnc}[1]{\textsf{#1}} % 
\newcommand{\lare}{\fnc{LaRe3D}}
\newcommand{\charc}{\fnc{CHAR}}

\newcommand{\DDsim}{DD{}}
\newcommand{\GTsim}{GT{}}

\newcommand{\vect}[1]{\boldsymbol{#1}}

\newcommand{\unit}[1]{\ensuremath{\, \mathrm{#1}}}

\definecolor{dylan}{rgb}{0, 0.5, 1.0}

%% file: 2025_DDBC_p3.bib
@ARTICLE{Afanasyev:2023,
       author = {{Afanasyev}, Andrey N. and {Fan}, Yuhong and {Kazachenko}, Maria D. and {Cheung}, Mark C.~M.},
        title = "{Hybrid Data-driven Magnetofrictional and Magnetohydrodynamic Simulations of an Eruptive Solar Active Region}",
      journal = {\apj},
         year = 2023,
        month = aug,
       volume = {952},
       number = {2},
          eid = {136},
        pages = {136},
          doi = {10.3847/1538-4357/acd7e9},
archivePrefix = {arXiv},
       eprint = {2306.05388},
 primaryClass = {astro-ph.SR},
       adsurl = {https://ui.adsabs.harvard.edu/abs/2023ApJ...952..136A}
}

@ARTICLE{Antiochos:1999,
       author = {{Antiochos}, S.~K. and {DeVore}, C.~R. and {Klimchuk}, J.~A.},
        title = "{A Model for Solar Coronal Mass Ejections}",
      journal = {\apj},
         year = 1999,
        month = jan,
       volume = {510},
       number = {1},
        pages = {485-493},
          doi = {10.1086/306563},
archivePrefix = {arXiv},
       eprint = {astro-ph/9807220},
 primaryClass = {astro-ph},
       adsurl = {https://ui.adsabs.harvard.edu/abs/1999ApJ...510..485A}
}

@ARTICLE{Arber:2001,
   author = {{Arber}, T.~D. and {Longbottom}, A.~W. and {Gerrard}, C.~L. and 
	{Milne}, A.~M.},
    title = "{A Staggered Grid, Lagrangian-Eulerian Remap Code for 3-D MHD Simulations}",
  journal = {Journal of Computational Physics},
     year = 2001,
    month = jul,
   volume = 171,
    pages = {151-181},
      doi = {10.1006/jcph.2001.6780},
   adsurl = {http://adsabs.harvard.edu/abs/2001JCoPh.171..151A}
}

@ARTICLE{Barnes:2024,
       author = {{Barnes}, Graham and {Hayashi}, Keiji and {Gilchrist}, S.~A.},
        title = "{Are Electric-field-driven Magnetohydrodynamic Simulations of the Solar Corona Sensitive to the Initial Condition?}",
      journal = {\apj},
         year = 2024,
        month = jan,
       volume = {960},
       number = {2},
          eid = {102},
        pages = {102},
          doi = {10.3847/1538-4357/ad10a7},
       adsurl = {https://ui.adsabs.harvard.edu/abs/2024ApJ...960..102B}
}

@ARTICLE{Chen:2023,
       author = {{Chen}, Feng and {Cheung}, Mark C.~M. and {Rempel}, Matthias and {Chintzoglou}, Georgios},
        title = "{Data-driven Radiative Magnetohydrodynamics Simulations with the MURaM Code}",
      journal = {\apj},
         year = 2023,
        month = jun,
       volume = {949},
       number = {2},
          eid = {118},
        pages = {118},
          doi = {10.3847/1538-4357/acc8c5},
archivePrefix = {arXiv},
       eprint = {2301.07621},
 primaryClass = {astro-ph.SR},
       adsurl = {https://ui.adsabs.harvard.edu/abs/2023ApJ...949..118C}
}

@ARTICLE{Chen:2025,
       author = {{Chen}, Feng},
        title = "{Data-driven Radiative Magnetohydrodynamics Simulations with the MURaM Code: the Emergence of Active Region 11158 and the X2.2 Flare}",
      journal = {arXiv e-prints},
         year = 2025,
        month = dec,
          eid = {arXiv:2512.01201},
        pages = {arXiv:2512.01201},
          doi = {10.48550/arXiv.2512.01201},
archivePrefix = {arXiv},
       eprint = {2512.01201},
 primaryClass = {astro-ph.SR},
       adsurl = {https://ui.adsabs.harvard.edu/abs/2025arXiv251201201C}
}

@ARTICLE{Cheung:2019,
       author = {{Cheung}, M.~C.~M. and {Rempel}, M. and {Chintzoglou}, G. and {Chen}, F. and {Testa}, P. and {Mart{\'\i}nez-Sykora}, J. and {Sainz Dalda}, A. and {DeRosa}, M.~L. and {Malanushenko}, A. and {Hansteen}, V. and {De Pontieu}, B. and {Carlsson}, M. and {Gudiksen}, B. and {McIntosh}, S.~W.},
        title = "{A comprehensive three-dimensional radiative magnetohydrodynamic simulation of a solar flare}",
      journal = {Nature Astronomy},
         year = 2019,
        month = nov,
       volume = {3},
        pages = {160-166},
          doi = {10.1038/s41550-018-0629-3},
       adsurl = {https://ui.adsabs.harvard.edu/abs/2019NatAs...3..160C}
}

@ARTICLE{Cimino:2016,
       author = {{Cimino}, A. and {Krause}, G. and {Elaskar}, S. and {Costa}, A.},
        title = "{Characteristic boundary conditions for magnetohydrodynamics: The Brio-Wu shocktube}",
      journal = {Computers and Fluids},
         year = 2016,
        month = mar,
       volume = {127},
        pages = {194-210},
          doi = {10.1016/j.compfluid.2016.01.001},
       adsurl = {https://ui.adsabs.harvard.edu/abs/2016CF....127..194C}
}

@ARTICLE{Courant:1928,
       author = {{Courant}, R. and {Friedrichs}, K. and {Lewy}, H.},
        title = "{{\"U}ber die partiellen Differenzengleichungen der mathematischen Physik}",
      journal = {Mathematische Annalen},
         year = 1928,
        month = jan,
       volume = {100},
        pages = {32-74},
          doi = {10.1007/BF01448839},
       adsurl = {https://ui.adsabs.harvard.edu/abs/1928MatAn.100...32C},
        note  = {Translated 1967, AEC Report NYO-7689, Phyllis Fox}
}

@BOOK{Courant:1953,
       author = {{Courant}, R. and {Hilbert}, D.},
        title = "{Methods of mathematical physics - Vol.1; Vol.2}",
         year = 1953,
        publisher = {Interscience Publishers, Inc.},
    place = {New York},
       adsurl = {https://ui.adsabs.harvard.edu/abs/1953mmp..book.....C}
}

@ARTICLE{Devore:2008,
       author = {{DeVore}, C. Richard and {Antiochos}, Spiro K.},
        title = "{Homologous Confined Filament Eruptions via Magnetic Breakout}",
      journal = {\apj},
         year = 2008,
        month = jun,
       volume = {680},
       number = {1},
        pages = {740-756},
          doi = {10.1086/588011},
       adsurl = {https://ui.adsabs.harvard.edu/abs/2008ApJ...680..740D}
}

@ARTICLE{Fan:2022,
       author = {{Fan}, Yuhong},
        title = "{An Improved Magnetohydrodynamic Simulation of the 2006 December 13 Coronal Mass Ejection of NOAA Active Region 10930}",
      journal = {\apj},
         year = 2022,
        month = dec,
       volume = {941},
       number = {1},
          eid = {61},
        pages = {61},
          doi = {10.3847/1538-4357/aca0ec},
archivePrefix = {arXiv},
       eprint = {2211.03736},
 primaryClass = {astro-ph.SR},
       adsurl = {https://ui.adsabs.harvard.edu/abs/2022ApJ...941...61F}}

@ARTICLE{Fan:2024,
       author = {{Fan}, Yuhong and {Kazachenko}, Maria D. and {Afanasyev}, Andrey N. and {Fisher}, George H.},
        title = "{A Data-driven Magnetohydrodynamic Simulation of the 2011 February 15 Coronal Mass Ejection from Active Region NOAA 11158}",
      journal = {\apj},
         year = 2024,
        month = nov,
       volume = {975},
       number = {2},
          eid = {206},
        pages = {206},
          doi = {10.3847/1538-4357/ad7f53},
archivePrefix = {arXiv},
       eprint = {2409.17507},
 primaryClass = {astro-ph.SR},
       adsurl = {https://ui.adsabs.harvard.edu/abs/2024ApJ...975..206F}
}

@ARTICLE{Guo:2024a,
       author = {{Guo}, J.~H. and {Ni}, Y.~W. and {Guo}, Y. and {Xia}, C. and {Schmieder}, B. and {Poedts}, S. and {Zhong}, Z. and {Zhou}, Y.~H. and {Yu}, F. and {Chen}, P.~F.},
        title = "{Data-driven Modeling of a Coronal Magnetic Flux Rope: From Birth to Death}",
      journal = {\apj},
         year = 2024,
        month = jan,
       volume = {961},
       number = {1},
          eid = {140},
        pages = {140},
          doi = {10.3847/1538-4357/ad088d},
archivePrefix = {arXiv},
       eprint = {2310.19617},
 primaryClass = {astro-ph.SR},
       adsurl = {https://ui.adsabs.harvard.edu/abs/2024ApJ...961..140G}
}

@ARTICLE{Guo:2024b,
       author = {{Guo}, Yang and {Guo}, Jinhan and {Ni}, Yiwei and {Xia}, Chun and {Zhong}, Ze and {Ding}, Mingde and {Chen}, Pengfei and {Keppens}, Rony},
        title = "{Magnetic flux rope models and data-driven magnetohydrodynamic simulations of solar eruptions}",
      journal = {Reviews of Modern Plasma Physics},
         year = 2024,
        month = dec,
       volume = {8},
       number = {1},
          eid = {29},
        pages = {29},
          doi = {10.1007/s41614-024-00167-2},
       adsurl = {https://ui.adsabs.harvard.edu/abs/2024RvMPP...8...29G}
}

@Article{Harris:2020,
 title         = {Array programming with {NumPy}},
 author        = {Charles R. Harris and K. Jarrod Millman and St{\'{e}}fan J.
                 van der Walt and Ralf Gommers and Pauli Virtanen and David
                 Cournapeau and Eric Wieser and Julian Taylor and Sebastian
                 Berg and Nathaniel J. Smith and Robert Kern and Matti Picus
                 and Stephan Hoyer and Marten H. van Kerkwijk and Matthew
                 Brett and Allan Haldane and Jaime Fern{\'{a}}ndez del
                 R{\'{i}}o and Mark Wiebe and Pearu Peterson and Pierre
                 G{\'{e}}rard-Marchant and Kevin Sheppard and Tyler Reddy and
                 Warren Weckesser and Hameer Abbasi and Christoph Gohlke and
                 Travis E. Oliphant},
 year          = {2020},
 month         = sep,
 journal       = {Nature},
 volume        = {585},
 number        = {7825},
 pages         = {357--362},
 doi           = {10.1038/s41586-020-2649-2},
 publisher     = {Springer Science and Business Media {LLC}},
 url           = {https://doi.org/10.1038/s41586-020-2649-2}
}

@PHDTHESIS{Harvey:1993,
       author = {{Harvey-Angle}, K.~L.},
        title = {Magnetic Bipoles on the Sun},
       school = {Utrecht University},
         year = 1993,
        month = jan,
       adsurl = {https://ui.adsabs.harvard.edu/abs/1993PhDT.......226H}
}

@ARTICLE{Harvey:1996,
       author = {{Harvey}, J.~W. and {Hill}, F. and {Hubbard}, R.~P. and {Kennedy}, J.~R. and {Leibacher}, J.~W. and {Pintar}, J.~A. and {Gilman}, P.~A. and {Noyes}, R.~W. and {Title}, A.~M. and {Toomre}, J. and {Ulrich}, R.~K. and {Bhatnagar}, A. and {Kennewell}, J.~A. and {Marquette}, W. and {Patron}, J. and {Saa}, O. and {Yasukawa}, E.},
        title = "{The Global Oscillation Network Group (GONG) Project}",
      journal = {Science},
         year = 1996,
        month = may,
       volume = {272},
       number = {5266},
        pages = {1284-1286},
          doi = {10.1126/science.272.5266.1284},
       adsurl = {https://ui.adsabs.harvard.edu/abs/1996Sci...272.1284H}
}

@ARTICLE{Hedstrom:1979,
   author = {{Hedstrom}, G.~W.},
    title = "{Nonreflecting Boundary Conditions for Nonlinear Hyperbolic Systems}",
  journal = {Journal of Computational Physics},
     year = 1979,
    month = feb,
   volume = 30,
    pages = {222-237},
      doi = {10.1016/0021-9991(79)90100-1},
   adsurl = {http://adsabs.harvard.edu/abs/1979JCoPh..30..222H}
}

@Article{Hunter:2007,
  Author    = {Hunter, J. D.},
  Title     = {Matplotlib: A 2D graphics environment},
  Journal   = {Computing in Science \& Engineering},
  Volume    = {9},
  Number    = {3},
  Pages     = {90--95},
  publisher = {IEEE COMPUTER SOC},
  doi       = {10.1109/MCSE.2007.55},
  year      = 2007
}

@ARTICLE{Inoue:2023,
       author = {{Inoue}, Satoshi and {Hayashi}, Keiji and {Miyoshi}, Takahiro},
        title = "{An Evolution and Eruption of the Coronal Magnetic Field through a Data-driven MHD Simulation}",
      journal = {\apj},
         year = 2023,
        month = mar,
       volume = {946},
       number = {1},
          eid = {46},
        pages = {46},
          doi = {10.3847/1538-4357/ac9eaa},
archivePrefix = {arXiv},
       eprint = {2210.07492},
 primaryClass = {astro-ph.SR},
       adsurl = {https://ui.adsabs.harvard.edu/abs/2023ApJ...946...46I}
}

@ARTICLE{Jiang:2021,
       author = {{Jiang}, Chaowei and {Bian}, Xinkai and {Sun}, Tingting and {Feng}, Xueshang},
        title = "{MHD Modeling of Solar Coronal Magnetic Evolution Driven by Photospheric Flow}",
      journal = {Frontiers in Physics},
         year = 2021,
        month = may,
       volume = {9},
          eid = {224},
        pages = {224},
          doi = {10.3389/fphy.2021.646750},
archivePrefix = {arXiv},
       eprint = {2104.07229},
 primaryClass = {astro-ph.SR},
       adsurl = {https://ui.adsabs.harvard.edu/abs/2021FrP.....9..224J}
}

@ARTICLE{Kee:2025,
       author = {{Kee}, N. Dylan and {Tarr}, Lucas A. and {Schuck}, Peter W. and {Linton}, Mark G. and {Leake}, James E.},
        title = "{Simulating the Photospheric to Coronal Plasma Using Magnetohydrodyanamic Characteristics. II. Reflections on Non-reflecting Boundary Conditions}",
      journal = {\apj},
         year = 2025,
        month = apr,
       volume = {983},
       number = {1},
          eid = {80},
        pages = {80},
          doi = {10.3847/1538-4357/adb132},
archivePrefix = {arXiv},
       eprint = {2501.17995},
 primaryClass = {astro-ph.SR},
       adsurl = {https://ui.adsabs.harvard.edu/abs/2025ApJ...983...80K}
}

@ARTICLE{Keller:2025,
       author = {{Keller}, Christoph U.},
        title = "{Data-driven Radiative Hydrodynamics Simulations of the Solar Photosphere Using Physics-informed Neural Networks: Proof of Concept}",
      journal = {\apj},
         year = 2025,
        month = aug,
       volume = {989},
       number = {1},
          eid = {92},
        pages = {92},
          doi = {10.3847/1538-4357/add6a9},
archivePrefix = {arXiv},
       eprint = {2505.04865},
 primaryClass = {astro-ph.SR},
       adsurl = {https://ui.adsabs.harvard.edu/abs/2025ApJ...989...92K}
}

@ARTICLE{Kusano:2020,
       author = {{Kusano}, Kanya and {Iju}, Tomoya and {Bamba}, Yumi and {Inoue}, Satoshi},
        title = "{A physics-based method that can predict imminent large solar flares}",
      journal = {Science},
         year = 2020,
        month = jul,
       volume = {369},
       number = {6503},
        pages = {587-591},
          doi = {10.1126/science.aaz2511},
       adsurl = {https://ui.adsabs.harvard.edu/abs/2020Sci...369..587K}
}

@ARTICLE{Leake:2017,
   author = {{Leake}, J.~E. and {Linton}, M.~G. and {Schuck}, P.~W.},
    title = "{Testing the Accuracy of Data-driven MHD Simulations of Active Region Evolution}",
  journal = {\apj},
archivePrefix = "arXiv",
   eprint = {1702.06808},
 primaryClass = "astro-ph.SR",
     year = 2017,
    month = apr,
   volume = 838,
      eid = {113},
    pages = {113},
      doi = {10.3847/1538-4357/aa6578},
   adsurl = {http://adsabs.harvard.edu/abs/2017ApJ...838..113L}
}

@ARTICLE{Leake:2022,
       author = {{Leake}, James E. and {Linton}, Mark G. and {Antiochos}, Spiro K.},
        title = "{The Role of Reconnection in the Onset of Solar Eruptions}",
      journal = {\apj},
         year = 2022,
        month = jul,
       volume = {934},
       number = {1},
          eid = {10},
        pages = {10},
          doi = {10.3847/1538-4357/ac74b7},
archivePrefix = {arXiv},
       eprint = {2205.12957},
 primaryClass = {astro-ph.SR},
       adsurl = {https://ui.adsabs.harvard.edu/abs/2022ApJ...934...10L}
}

@ARTICLE{Linton:1996,
       author = {{Linton}, M.~G. and {Longcope}, D.~W. and {Fisher}, G.~H.},
        title = "{The Helical Kink Instability of Isolated, Twisted Magnetic Flux Tubes}",
      journal = {\apj},
         year = 1996,
        month = oct,
       volume = {469},
        pages = {954},
          doi = {10.1086/177842},
       adsurl = {https://ui.adsabs.harvard.edu/abs/1996ApJ...469..954L}
}

@ARTICLE{Liu:2024,
       author = {{Liu}, Zhi-Peng and {Jiang}, Chao-Wei and {Bian}, Xin-Kai and {Liu}, Qing-Jun and {Zou}, Peng and {Feng}, Xue-Shang},
        title = "{A New Approach of Data-driven Simulation and its Application to Solar Active Region 12673}",
      journal = {Research in Astronomy and Astrophysics},
         year = 2024,
        month = dec,
       volume = {24},
       number = {12},
          eid = {125005},
        pages = {125005},
          doi = {10.1088/1674-4527/ad862b},
archivePrefix = {arXiv},
       eprint = {2410.09433},
 primaryClass = {astro-ph.SR},
       adsurl = {https://ui.adsabs.harvard.edu/abs/2024RAA....24l5005L}
}

@book{Morse:1953a,
  address = {New York},
  author = {{Morse}, P.~M. and {Feshbach}, H. },
  publisher = {McGraw-Hill},
  title = {Methods of theoretical physics, Part I},
  year = 1953
}

@ARTICLE{Norton:2017,
       author = {{Norton}, A.~A. and {Jones}, E.~H. and {Linton}, M.~G. and {Leake}, J.~E.},
        title = "{Magnetic Flux Emergence and Decay Rates for Preceder and Follower Sunspots Observed with HMI}",
      journal = {\apj},
         year = 2017,
        month = jun,
       volume = {842},
       number = {1},
          eid = {3},
        pages = {3},
          doi = {10.3847/1538-4357/aa7052},
archivePrefix = {arXiv},
       eprint = {1705.02053},
 primaryClass = {astro-ph.SR},
       adsurl = {https://ui.adsabs.harvard.edu/abs/2017ApJ...842....3N}
}

@ARTICLE{Pesnell:2012,
   author = {{Pesnell}, W.~D. and {Thompson}, B.~J. and {Chamberlin}, P.~C.
	},
    title = "{The Solar Dynamics Observatory (SDO)}",
  journal = {\solphys},
     year = 2012,
    month = jan,
   volume = 275,
    pages = {3-15},
      doi = {10.1007/s11207-011-9841-3},
   adsurl = {http://adsabs.harvard.edu/abs/2012SoPh..275....3P}
}

@BOOK{Priest:2014,
       author = {{Priest}, Eric},
        title = "{Magnetohydrodynamics of the Sun}",
         year = 2014,
          doi = {10.1017/CBO9781139020732},
    publisher = {Cambridge University Press},
       place  = {Cambridge},
       adsurl = {https://ui.adsabs.harvard.edu/abs/2014masu.book.....P}
}

@ARTICLE{Rempel:2014,
       author = {{Rempel}, M.},
        title = "{Numerical Simulations of Quiet Sun Magnetism: On the Contribution from a Small-scale Dynamo}",
      journal = {\apj},
         year = 2014,
        month = jul,
       volume = {789},
       number = {2},
          eid = {132},
        pages = {132},
          doi = {10.1088/0004-637X/789/2/132},
archivePrefix = {arXiv},
       eprint = {1405.6814},
 primaryClass = {astro-ph.SR},
       adsurl = {https://ui.adsabs.harvard.edu/abs/2014ApJ...789..132R}
}

@ARTICLE{SainzDalda:2019,
       author = {{Sainz Dalda}, Alberto and {de la Cruz Rodr{\'\i}guez}, Jaime and {De Pontieu}, Bart and {Go{\v{s}}i{\'c}}, Milan},
        title = "{Recovering Thermodynamics from Spectral Profiles observed by IRIS: A Machine and Deep Learning Approach}",
      journal = {\apjl},
         year = 2019,
        month = apr,
       volume = {875},
       number = {2},
          eid = {L18},
        pages = {L18},
          doi = {10.3847/2041-8213/ab15d9},
archivePrefix = {arXiv},
       eprint = {1904.08390},
 primaryClass = {astro-ph.SR},
       adsurl = {https://ui.adsabs.harvard.edu/abs/2019ApJ...875L..18S}
}

@ARTICLE{Scherrer:1995,
       author = {{Scherrer}, P.~H. and {Bogart}, R.~S. and {Bush}, R.~I. and {Hoeksema}, J.~T. and {Kosovichev}, A.~G. and {Schou}, J. and {Rosenberg}, W. and {Springer}, L. and {Tarbell}, T.~D. and {Title}, A. and {Wolfson}, C.~J. and {Zayer}, I. and {MDI Engineering Team}},
        title = "{The Solar Oscillations Investigation - Michelson Doppler Imager}",
      journal = {\solphys},
         year = 1995,
        month = dec,
       volume = {162},
       number = {1-2},
        pages = {129-188},
          doi = {10.1007/BF00733429},
       adsurl = {https://ui.adsabs.harvard.edu/abs/1995SoPh..162..129S}
}

@ARTICLE{Scherrer:2012,
   author = {{Scherrer}, P.~H. and {Schou}, J. and {Bush}, R.~I. and {Kosovichev}, A.~G. and 
	{Bogart}, R.~S. and {Hoeksema}, J.~T. and {Liu}, Y. and {Duvall}, T.~L. and 
	{Zhao}, J. and {Title}, A.~M. and {Schrijver}, C.~J. and {Tarbell}, T.~D. and 
	{Tomczyk}, S.},
    title = "{The Helioseismic and Magnetic Imager (HMI) Investigation for the Solar Dynamics Observatory (SDO)}",
  journal = {\solphys},
     year = 2012,
    month = jan,
   volume = 275,
    pages = {207-227},
      doi = {10.1007/s11207-011-9834-2},
   adsurl = {http://cdsads.u-strasbg.fr/abs/2012SoPh..275..207S}
}

@ARTICLE{Schmieder:2024,
       author = {{Schmieder}, Brigitte and {Guo}, Jinhan and {Poedts}, Stefaan},
        title = "{Recent advances in solar data-driven MHD simulations of the formation and evolution of CME flux ropes}",
      journal = {Reviews of Modern Plasma Physics},
         year = 2024,
        month = aug,
       volume = {8},
       number = {1},
          eid = {27},
        pages = {27},
          doi = {10.1007/s41614-024-00166-3},
archivePrefix = {arXiv},
       eprint = {2408.06595},
 primaryClass = {astro-ph.SR},
       adsurl = {https://ui.adsabs.harvard.edu/abs/2024RvMPP...8...27S}
}

@ARTICLE{Tarr:2024,
       author = {{Tarr}, Lucas A. and {Kee}, N. Dylan and {Linton}, Mark G. and {Schuck}, Peter W. and {Leake}, James E.},
        title = "{Simulating the Photospheric to Coronal Plasma Using Magnetohydrodynamic Characteristics. I. Data-driven Boundary Conditions}",
      journal = {\apjs},
         year = 2024,
        month = feb,
       volume = {270},
       number = {2},
          eid = {30},
        pages = {30},
          doi = {10.3847/1538-4365/ad0e0c},
archivePrefix = {arXiv},
       eprint = {2311.02281},
 primaryClass = {astro-ph.SR},
       adsurl = {https://ui.adsabs.harvard.edu/abs/2024ApJS..270...30T}
}

@ARTICLE{Thompson:1987a,
   author = {{Thompson}, K.~W.},
    title = "{Time dependent boundary conditions for hyperbolic systems}",
  journal = {Journal of Computational Physics},
     year = 1987,
    month = jan,
   volume = 68,
    pages = {1-24},
      doi = {10.1016/0021-9991(87)90041-6},
   adsurl = {http://adsabs.harvard.edu/abs/1987JCoPh..68....1T}
}

@ARTICLE{Thompson:1990,
   author = {{Thompson}, K.~W.},
    title = "{Time-dependent boundary conditions for hyperbolic systems. II}",
  journal = {Journal of Computational Physics},
     year = 1990,
    month = aug,
   volume = 89,
    pages = {439-461},
      doi = {10.1016/0021-9991(90)90152-Q},
   adsurl = {http://adsabs.harvard.edu/abs/1990JCoPh..89..439T}
}

@ARTICLE{Vogler:2005,
       author = {{V{\"o}gler}, A. and {Shelyag}, S. and {Sch{\"u}ssler}, M. and {Cattaneo}, F. and {Emonet}, T. and {Linde}, T.},
        title = "{Simulations of magneto-convection in the solar photosphere.  Equations, methods, and results of the MURaM code}",
      journal = {\aap},
         year = 2005,
        month = jan,
       volume = {429},
        pages = {335-351},
          doi = {10.1051/0004-6361:20041507},
       adsurl = {https://ui.adsabs.harvard.edu/abs/2005A&A...429..335V}
}

@ARTICLE{Wagner:2024,
       author = {{Wagner}, A. and {Price}, D.~J. and {Bourgeois}, S. and {Daei}, F. and {Pomoell}, J. and {Poedts}, S. and {Kumari}, A. and {Barata}, T. and {Erd{\'e}lyi}, R. and {Kilpua}, E.~K.~J.},
        title = "{The effect of data-driving and relaxation models on magnetic flux rope evolution and stability}",
      journal = {\aap},
         year = 2024,
        month = dec,
       volume = {692},
          eid = {A74},
        pages = {A74},
          doi = {10.1051/0004-6361/202450577},
archivePrefix = {arXiv},
       eprint = {2410.18672},
 primaryClass = {astro-ph.SR},
       adsurl = {https://ui.adsabs.harvard.edu/abs/2024A&A...692A..74W}
}
